\documentclass[acmsmall]{acmart}

\usepackage{array,etoolbox}
\preto\tabular{\setcounter{magicrownumbers}{0}}
\newcounter{magicrownumbers}

\usepackage{multirow}
\usepackage{amsmath}
\usepackage{caption}
\usepackage{subcaption}
\usepackage{pdfpages}

\newcommand{\pov}{point-of-view}
\newcommand{\npov}{Neutral Point of View (NPOV) }

\newcommand{\clarif}{clarification }

\AtBeginDocument{%
  \providecommand\BibTeX{{%
    \normalfont B\kern-0.5em{\scshape i\kern-0.25em b}\kern-0.8em\TeX}}}

\setcopyright{acmlicensed}
\acmJournal{PACMHCI}
\acmYear{2021} \acmVolume{5} \acmNumber{CSCW2} \acmArticle{359} \acmMonth{10} \acmPrice{15.00}\acmDOI{10.1145/3479503}

\begin{document}

%%
%% The "title" command has an optional parameter,
%% allowing the author to define a "short title" to be used in page headers.
\title{Automatically Labeling Low Quality Content on Wikipedia by Leveraging Patterns in Editing Behavior}
%\title{Automatically Building Content Flaw Detection Models By Learning From Expert Behaviors}

%%
%% The "author" command and its associated commands are used to define
%% the authors and their affiliations.
%% Of note is the shared affiliation of the first two authors, and the
%% "authornote" and "authornotemark" commands
%% used to denote shared contribution to the research.
\author{Sumit Asthana}
\email{asumit@umich.edu}
\affiliation{%
  \institution{University of Michigan}
  %\streetaddress{P.O. Box 1212}
  \city{Ann Arbor}
  \state{Michigan}
  \country{USA}
  %\postcode{43017-6221}
}

\author{Sabrina Tobar Thommel}
\email{sabtt@umich.edu}
\affiliation{%
  \institution{University of Michigan}
  %\streetaddress{P.O. Box 1212}
  \city{Ann Arbor}
  \state{Michigan}
  \country{USA}
  %\postcode{43017-6221}
}

\author{Aaron Lee Halfaker}
\email{ahalfaker@microsoft.com}
\affiliation{%
  \institution{Microsoft}
  %\streetaddress{P.O. Box 1212}
  \city{Seattle}
  \state{Washington}
  \country{USA}
  %\postcode{43017-6221}
}

\author{Nikola Banovic}
\email{nbanovic@umich.edu}
\affiliation{%
  \institution{University of Michigan}
  %\streetaddress{P.O. Box 1212}
  \city{Ann Arbor}
  \state{Michigan}
  \country{USA}
  %\postcode{43017-6221}
}

%%
%% By default, the full list of authors will be used in the page
%% headers. Often, this list is too long, and will overlap
%% other information printed in the page headers. This command allows
%% the author to define a more concise list
%% of authors' names for this purpose.
\renewcommand{\shortauthors}{Asthana et al.}

% References to add
% NPOV case study - https://twitter.com/snaglee2401/status/1362493353007398915
% Reddit moderation - https://dl.acm.org/doi/abs/10.1145/3338243

\begin{abstract}
  Wikipedia articles aim to be definitive sources of encyclopedic content. Yet, only 0.6\% of Wikipedia articles have high quality according to its quality scale due to insufficient number of Wikipedia editors and enormous number of articles. Supervised Machine Learning (ML) quality improvement approaches that can automatically identify and fix content issues rely on manual labels of individual Wikipedia sentence quality. However, current labeling approaches are tedious and produce noisy labels. Here, we propose an automated labeling approach that identifies the semantic category (e.g., adding citations, clarifications) of historic Wikipedia edits and uses the modified sentences prior to the edit as examples that require that semantic improvement. Highest-rated article sentences are examples that no longer need semantic improvements. We show that training existing sentence quality classification algorithms on our labels improves their performance compared to training them on existing labels. Our work shows that editing behaviors of Wikipedia editors provide better labels than labels generated by crowdworkers who lack the context to make judgments that the editors would agree with.
\end{abstract}

%%
%% The code below is generated by the tool at http://dl.acm.org/ccs.cfm.
%% Please copy and paste the code instead of the example below.
%%
\begin{CCSXML}
<ccs2012>
<concept>
<concept_id>10003120.10003130.10003131.10003270</concept_id>
<concept_desc>Human-centered computing~Social recommendation</concept_desc>
<concept_significance>500</concept_significance>
</concept>
<concept>
<concept_id>10003120.10003130.10003131.10003570</concept_id>
<concept_desc>Human-centered computing~Computer supported cooperative work</concept_desc>
<concept_significance>500</concept_significance>
</concept>
<concept>
<concept_id>10003120.10003130.10011762</concept_id>
<concept_desc>Human-centered computing~Empirical studies in collaborative and social computing</concept_desc>
<concept_significance>500</concept_significance>
</concept>
<concept>
<concept_id>10003120.10003130.10003233.10003301</concept_id>
<concept_desc>Human-centered computing~Wikis</concept_desc>
<concept_significance>300</concept_significance>
</concept>
<concept>
<concept_id>10003120.10003130.10003233.10010922</concept_id>
<concept_desc>Human-centered computing~Social tagging systems</concept_desc>
<concept_significance>300</concept_significance>
</concept>
</ccs2012>
\end{CCSXML}

\ccsdesc[500]{Human-centered computing~Social recommendation}
\ccsdesc[500]{Human-centered computing~Computer supported cooperative work}
\ccsdesc[500]{Human-centered computing~Empirical studies in collaborative and social computing}
\ccsdesc[300]{Human-centered computing~Wikis}
\ccsdesc[300]{Human-centered computing~Social tagging systems}
\keywords{Wikipedia, Labeling, Machine Learning.}

%%
%% This command processes the author and affiliation and title
%% information and builds the first part of the formatted document.
\maketitle
\section{Introduction}
Wikipedia \cite{Wikipedia2020}, an online encyclopedia, aims to be the ultimate source of encyclopedic knowledge by achieving a high quality for all its articles. High quality articles are definitive source of knowledge on the topic and serve the purpose of providing information to Wikipedia readers in a concise manner, without causing confusion and wasting time~\cite{DBLP:journals/jmis/WangS96}. Thus, Wikipedia editors have defined a comprehensive content assessment criteria, called the WP1.0 Article Quality Assessment scale \cite{WikipediaContentAssessment2020} to grade article quality on a scale from the most basic "stub" (articles with basic information about the topic, without proper citations and Wikipedia-defined structure) to the exemplary "Featured Articles" (well-written and well-structured, comprehensive and properly cited articles).

\begin{figure}[t]
    \centering
    \includegraphics[width=\textwidth]{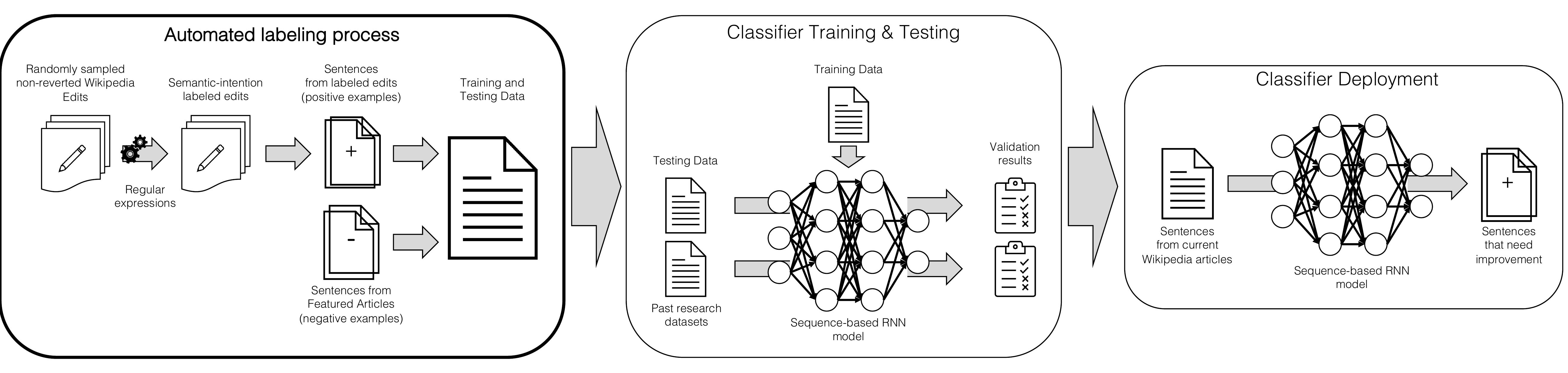}
    \caption{Our pipeline for labeling low-quality sentences on Wikipedia. We start with our automated labeling approach (left), where we obtain a large corpus of historic Wikipedia sentence edits, and label their semantic intent using programmatic rules. We extract positive sentences from relevant semantic edits and negative sentences from Featured Articles. We then use our labels to train existing Machine Learning models, and test them by comparing with labeling approaches  from past research (middle). Existing models trained on our labels can then be deployed to automatically detect Wikipedia sentences that require improvement (right).}
    \label{fig:piepline}
\end{figure}

Article maintenance, as opposed to creating new articles and content, has become a significant portion of what Wikipedia editors do \cite{DBLP:conf/chi/KitturSPC07}.
Currently, editors rate article quality and identify and make required improvements manually, which is taxing and time-consuming. Being a collaborative editing platform, articles are in a constant state of churn and current assessments are quickly outdated because articles will have been modified by others. For the limited number of experienced editors on Wikipedia, performing such assessments across a set of 6.5 million Wikipedia articles is a huge bottleneck~\cite{stvilia2008information}; currently only about 7,000 of articles have "Featured Article" status and only about 33,000 have the second best "Good Article" status \cite{WikipediaContentAssessment2020}.

With continuously declining number of editors on Wikipedia \cite{DBLP:conf/wikis/SuhCCP09}, automating quality assessment tasks could reduce the workload of remaining editors. Supervised Machine Learning (ML) has already automated tasks like vandalism detection \cite{DBLP:conf/wikis/Halfaker17} and overall article quality prediction \cite{DBLP:conf/wikis/Warncke-WangCR13}. Such ML approaches require labeled sets of examples of Wikipedia content that requires improvement (positive examples) and content that does not (negative examples). One of the main reasons for the success of those existing ML approaches \cite{DBLP:conf/wikis/Halfaker17, DBLP:conf/wikis/Warncke-WangCR13} (both have been deployed to Wikipedia) is the relative ease of obtaining labels either because they are visually salient (e.g., in case of vandalism) or already part of existing practices (e.g., editors manually record article quality on talk pages of Wikipedia articles as part of existing article assessment).

However, automating other quality assessment tasks (e.g., identifying sentences that require citation, sentences with non-neutral point of view, sentences that require clarification) requires labels at the Wikipedia sentence level which makes automating such tasks difficult. Wikipedia editors rarely manually flag outstanding Wikipedia sentence quality issues as part of their editing process~\cite{DBLP:conf/sigir/AnderkaSL12}. Even existing crowdsourcing-based labeling method \cite{DBLP:conf/wsdm/HubeF19, DBLP:conf/emnlp/YangHKH17,DBLP:conf/www/RediFMT19} could produce noisy Wikipedia sentence quality labels, especially when crowdworkers, who are not domain experts, lack knowledge about Wikipedia policies on content quality~\cite{Kittur2008, Forte2018, DBLP:conf/fat/GeigerYYDQTH20, kairam2016parting}.

Here, we present a method for automatically labeling Wikipedia sentence quality across improvement categories directly from past Wikipedia editors' editing behavior to enable automated detection of sentences that need quality improvements (Figure \ref{fig:piepline}). To label positive examples (sentences that need improvements), we implemented Wikipedia core content principles guidelines~\cite{WikipediaCore2020} as syntax-based rules to capture the meaning or intent of a historic Wikipedia edit (i.e., the smallest recorded unit of change in a Wikipedia article, paragraph, or sentence, such as added citations, removed bias, clarification) for each quality category we want to classify (e.g., needs citation, needs bias-removal, or needs clarification). Each historic edit then indicates that the edited sentence needed that particular improvement resulting in a positive example. We follow Redi et al's.~\cite{DBLP:conf/www/RediFMT19} approach and label all sentences in featured articles as negative examples (sentences that do not need improvements). 

%\includepdf[landscape=true]{images/pipeline.pdf}

% No longer
%Our key insight is - \textit{while it is not feasible to ask Wikipedia editors to label enough statements with their quality to be able to train model to automatically detect which statements need improvement, it is easy to evaluate the quality prediction of these models and modify the rules to re-sample the training data as per the feedback.}

To illustrate our approach, we built three Wikipedia sentence quality detection pipelines (including corresponding rules) for three Wikipedia quality improvement categories: 1) citations (adding or modifying references and citations for verifiability), 2) \npov edits (rewriting using encyclopedic, neutral tone; removing bias), and 3) clarifications (specifying or explaining an existing fact or meaning by example or discussion without adding new information)~\cite{DBLP:conf/emnlp/YangHKH17}. We first evaluated our rules in a user study with nine Wikipedia editors, in which they manually labeled improvement category of 434 historic Wikipedia edits. We then compared the outputs of our rules with the ground truth and showed that our rules could effectively extract positive examples. Our results also revealed high ambiguity amongst participants' manual labels, which further underline the importance of our automated rule-based approach.

We then validated the usefulness of our automated labeling approach by comparing the performance of \textit{existing} deep learning models~\cite{DBLP:journals/corr/BahdanauCB14} trained using existing, baseline labeling approaches (e.g., implicit labeling \cite{DBLP:conf/www/RediFMT19}, crowdsourcing \cite{DBLP:conf/wsdm/HubeF19}) and our automatically extracted labels. Our results showed that existing models trained using our automatic labeling method achieved 29\% and 22\% improvement in F1-score for citations and NPOV respectively than the same models trained on data labeled using existing approaches.

Our work provides further evidence that the edits produced by Wikipedians working in their context provide better signal for supporting their work than labels generated by crowdworkers who lack the context to make judgments about sentence quality that Wikipedians would agree with. Learning from implicit editing behavior of Wikipedia editors allowed us to produce labels that capture the nuances of Wikipedia quality policies~\cite{keegan2017evolution}. Our work has implications for the growth of collaborative content spaces where different people come together to curate content adhering to the standards and purpose of the space~\cite{morgan2018welcome}. With Wikipedia behavior policies becoming more decentralized~\cite{forte2009decentralization}, our strategy of learning from implicit behaviors has additional relevance of enforcing norms that may only be enacted through reading and observing policies up  until now.

%While a diverse set of people curating the content has its benefits, people bring their own biases to the table. Thus, information quality is an important dimension of such spaces to make sure the content follows certain standards and guidelines that make it usable in the intended space \cite{lukyanenko2014iq}. Assessing information quality, let alone improving the content is often a manual intensive task. Automating ways to assess and improve information quality using Machine Learning(ML) requires labels of quality on the content. These labels are hard to obtain because of unavailability of experts for intensive content quality annotation tasks. Obtaining implicit labels or crowdsourcing are possible alternatives for getting such quality labels. Crowdworkers are not domain experts and hence quality prediction models learnt from crowdlabels are often at best not usable for effective content flaw detections \cite{mozetivc2016multilingual}. Implicit labels are a viable alternative but they should be extracted with care. By taking Wikipedia as an example, we propose a method to automatically build content flaw detection models by learning from expert editors behaviors.

\section{Collaborative Platform Content Quality Labeling Challenges}
\label{sec:related}
Research has given considerable attention to improving and maintaining good quality of user generated content on collaborative platforms. Such research has explored both assisted and automated editing tools~\cite{geiger2010work} for content creation \& recommendation~\cite{wulczyn2016growing}, vandalism detection~\cite{DBLP:conf/wikis/Halfaker17, DBLP:conf/wikis/GeigerH13}, and content regulation (e.g., content that violates platform policy) using both programmatic rules~\cite{jhaver2019human, DBLP:journals/pacmhci/GeigerH17, DBLP:conf/group/PriedhorskyCLPTR07,fiesler2018reddit} and Machine Learning-based methods~\cite{chandrasekharan2019crossmod}. A subset of such research focuses specifically on automatically detecting content quality (e.g.,~\cite{DBLP:conf/wikis/Warncke-WangCR13, DBLP:conf/jcdl/DalipGCC09}) and bringing the issues to the attention of the community.
%Even on Wikipedia, rule based approaches have shown promise in identifying vandalism-related edit reverts ~\cite{DBLP:conf/group/PriedhorskyCLPTR07} and distinguishing between conflict and non-conflict edit revert activity ~\cite{DBLP:journals/pacmhci/GeigerH17}.

Such automated efforts have been possible in part because of the availability of quality labels for such tasks.
For example, a small subset of visually-salient, hand-labeled examples are sufficient for simple ML models to identify vandalism with high accuracy~\cite{DBLP:conf/wikis/GeigerH13}. Also, training existing article quality models~\cite{DBLP:conf/wikis/Warncke-WangCR13} involves using existing quality labels for over 6.5 million articles that Wikipedians have generated when manually rating article quality as part of existing processes.

Unfortunately, that kind of automated assistance to editors does not easily extend to other collaborative editing tasks because labels for other quality assessment and improvement tasks are not immediately available.
For example, although Wikipedia encourages its editors to manually flag outstanding content issues with cleanup templates markup (e.g., marking a sentence with \textit{\{citation needed\}} template~\cite{WikipediaTemplateCleanup2020}) or label their Wikipedia edits with a free-form edit intent summary (e.g., \pov), their usage is not standardised and only few Wikipedia edits or Wikipedia sentences that need improvement actually have them~\cite{DBLP:conf/sigir/AnderkaSL12}.

Existing attempts to supplement such labels \textit{via} crowdsourcing ~\cite{DBLP:conf/wsdm/HubeF19, DBLP:conf/emnlp/YangHKH17,DBLP:conf/www/RediFMT19} produce too few labels when using Wikipedia editors as labelers or produce noisy labels when using crowdworkers who are not Wikipedia editors. Such non-editor crowdworkers do not always provide reliable judgments on what content needs improvement~\cite{ daniel2018quality}, often due to their lack of knowledge about the nuances of Wikipedia policies~\cite{Kittur2008, Forte2018}.

Although crowdsourcing has been used in the past~\cite{DBLP:conf/sigir/Potthast10} to successfully label examples of vandalism, it is important to note that annotating vandalism is simpler than examples related to concepts like the need for citations, neutrality of point-of-view, and clarifications, since the concept of vandalism could be commonly shared between lay Web users and Wikipedians. In the absence of a widely accepted clear standard of categorization of Wikipedia sentence quality, most of the other tasks that editors perform are hard to label.

% REVIEW THIS
To get around explicitly asking editors or crowdworkers to label the quality of Wikipedia sentences, existing research ~\cite{DBLP:conf/www/RediFMT19} has attempted to obtain labels implicitly.
%by propagating article quality label to all sentences in the article.
Redi et al. ~\cite{DBLP:conf/www/RediFMT19} showed that citation labels are easy to obtain because presence/absence of citations in "Featured Articles" acts as an implicit label that sentences with citations needed them and those without did not. While such implicit labeling strategy can be used to label negative examples across semantic improvement categories, they cannot extract positive examples of needed improvements for categories, such as neutrality of point-of-view or clarification.

Recently, Yang et al. ~\cite{DBLP:conf/emnlp/YangHKH17} created a taxonomy of Wikipedia edits based on the semantic intention behind the edit to build a classifier to automatically categorize the semantic intent of Wikipedia edits. This taxonomy comprehensively covers the tasks Wikipedians do, ranging from fighting vandalism and copy-editing to making content clarifications and simplifications. Existing research~\cite{DBLP:conf/emnlp/YangHKH17, ruprechter2020relating} has used this taxonomy to automatically classify article quality using ML-based approaches, but not individual Wikipedia sentence quality. Thus, such existing approaches do not pinpoint which specific parts of a Wikipedia article need improvement.
Unfortunately, also, it is not immediately obvious how to adapt such existing article quality labeling methods to the problem of automatically labeling Wikipedia sentence quality using the taxonomy above.
 
%  \subsection{DRAFT}
%  manual labor of moderators \cite{rader2015understanding,danieldylan}. Cross community learning - learning from one site and applying it to another \cite{chen2014laborers}. The evolution and consequences of peer producing Wikipedia's rules\cite{keegan2017evolution}. "Relating Wikipedia article quality to link structure and edit behavior"\cite{ruprechter2020relating}. "Growing Wikipedia across languages via recommendation" \cite{wulczyn2016growing}

% \subsection{Technical enforcement of normative behavior}
% Automated approaches for technical enforcement of norms are already being used in Wikipedia\cite{lovink2012critical}. Decentralized governance in Wikipedia\cite{forte2009decentralization}.

\section{Method for Automatically Labeling Low Quality Content}
\label{sec:method}
To build a pipeline for automatic detection of Wikipedia sentences that need quality improvements (Figure~\ref{fig:piepline}), we need examples of sentences that require improvement (i.e., positive examples) and examples of sentences that do not need any further improvement (i.e., negative examples). We first focus on extracting positive examples, which is the main contribution of our work. We then use an existing method~\cite{DBLP:conf/www/RediFMT19} for extracting negative examples, which assumes sentences from Featured Articles do not need further improvements (we only briefly summarize it in this section). We can then train different ML models for each semantic category on such labeled sentences, and use the classifier to classify if previously unseen Wikipedia sentences (e.g., newly added or edited Wikipedia sentences that we have not trained on) need a particular kind of semantic improvement.

\subsection{Identifying Semantic Intents in Wikipedia Edits to Extract Positive Examples}
Here, we leverage traces of Wikipedia editors' collaborative editing behaviors to learn which Wikipedia edits are attempts by the editors to improve quality of Wikipedia sentences. Our approach has three benefits: 1) collaborative editing behavior data is readily available because Wikipedia automatically logs all edits as part of its editing process, 2) editing is common across all Wikipedia languages, hence provides for a common approach for detecting quality issues for Wikipedia in all languages, and 3) Wikipedia already provides a definition for categorization of edits based on their semantic intent~\cite{WikipediaCore2020} (e.g., adding citations, removing point of view, clarifying Wikipedia sentences). 

Our insight is that it is easier to identify semantic intent of edits given their syntax compared to identifying quality issues directly in free form natural language sentences. Unlike existing work~\cite{DBLP:conf/emnlp/YangHKH17} that attempted to identify the semantic intent across multiple Wikipedia paragraphs (each composed of multiple Wikipedia sentences), we identify semantic improvements at Wikipedia sentence level to pinpoint which sentences need improvement. This also minimizes the noise associated with propagating weak, paragraph-level labels to other sentences in the paragraph that do not need such specific improvement.

\begin{figure}[h]
    \includegraphics[trim=20 60 20 60,clip,width=\textwidth]{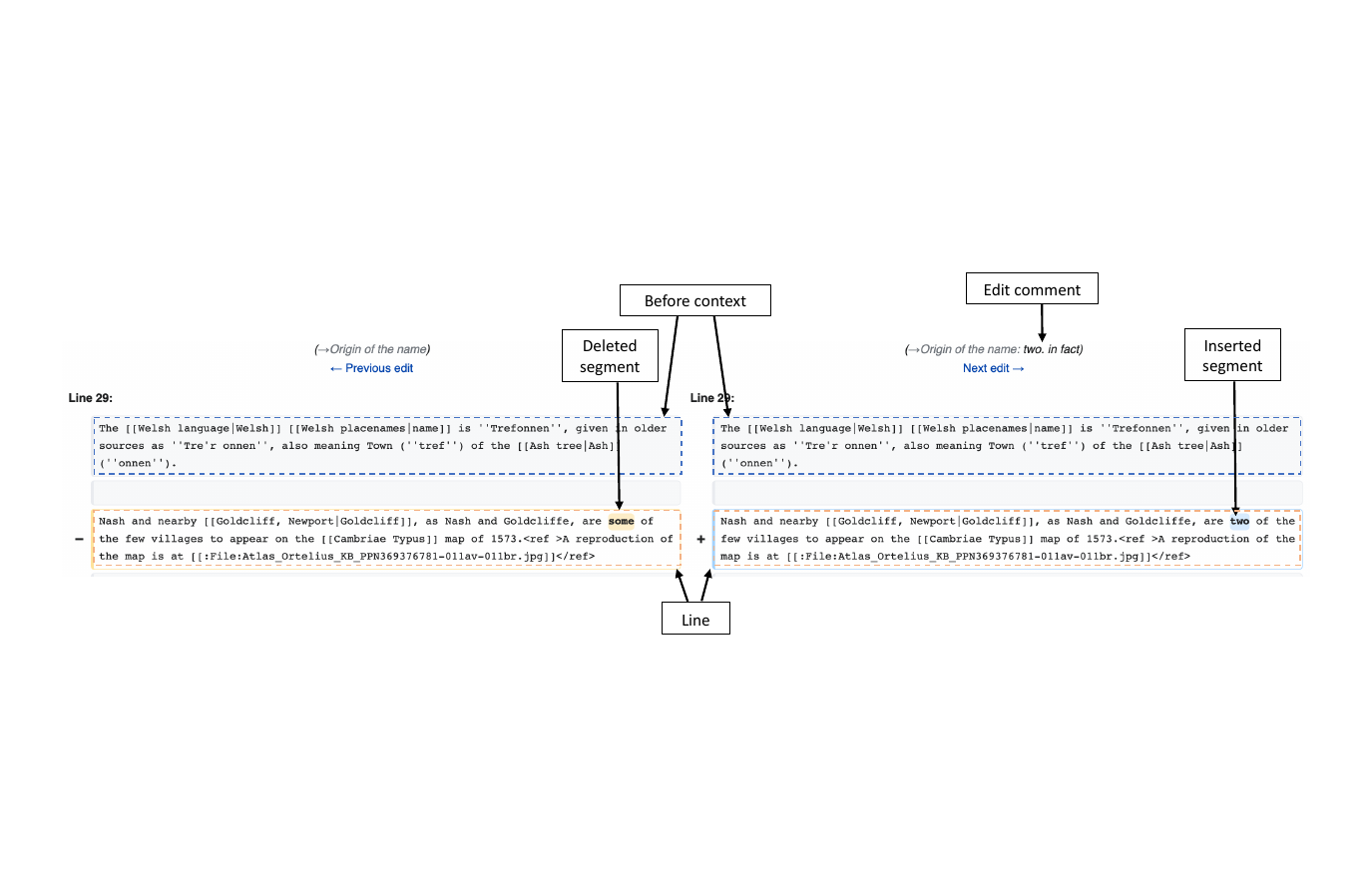}
    \caption{An example of an edit diff showing two segments - inserted and deleted. Before context is shown with blue dashed line. Orange dashed line outlines one full line.}
    \label{fig:segment-example}
\end{figure}

Thus, we start with automatically identifying the semantic intent of different Wikipedia edits, which we later use to indicate that the Wikipedia sentence prior to being edited required that semantic improvement (e.g., that it required adding citations, removing point of view, or clarifying). Each time an editor edits an article, Wikipedia represents all of the changes the editor made at that time as an \textit{edit diff} (Figure \ref{fig:segment-example}), which consists of Wikipedia content (e.g., section headings, paragraphs, sentences) before editor changes, the same content after editor changes, and an optional comment from the editor that contains their explanation for the edit. Note that an \textit{edit diff} can span multiple Wikipedia article sections and paragraphs. Wikipedia splits each \textit{edit diff} into one or more \textit{lines}, which indicate exact content that the editor changed and the surrounding content (i.e., \textit{context}). Each changed \textit{line} contains one or more \textit{segments}, each representing a continuous unit of change (i.e., there is no content within the segment that is unchanged).

Since we want to label individual Wikipedia sentences within an \textit{edit diff}, we only extract sentences from the \textit{edit diff} that contain at least one \textit{segment}. To extract positive examples, we test each changed sentence against a set of rules, which translate natural language description of semantic edits categorization~\cite{WikipediaCore2020} to computer code. Here, we illustrate our method on three semantic categories: 1) \textit{Citations} ("add or modify references and citations; remove unverified text"), 2) \textit{Point-of-View (POV)} ("rewrite using encyclopedic, neutral tone; remove bias; apply due weight"), 3) \textit{Clarifications} ("specify or explain an existing fact or meaning by example or discussion without adding new information"). Table \ref{tab:rules-cat} specifies the rules for the three semantic categories. A specific semantic category (e.g., point-of-view) is applied only when all the rules under the given category satisfy. Table \ref{tab:rules-regex} shows regular expressions that we implemented for each rule in Table \ref{tab:rules-cat}.

We started with a set of rules that implemented our own subjective interpretation of the semantic intent categories. We then iterated on our rules by evaluating them on single-line \textit{edit diffs} from the \textit{edittypes} dataset~\cite{DBLP:conf/emnlp/YangHKH17} and comparing the outputs of our rules with the \textit{edit diffs} labels from that dataset to tune the parameters of our rules (e.g., to determine the value of inserted\_length\_words parameter in Table \ref{tab:rules-cat} which represents the number of inserted words in a \textit{segment}).

Note that the existing dataset~\cite{DBLP:conf/emnlp/YangHKH17} contains crowdsourced labels, which could be noisy. We therefore used the dataset as a reference, and not as ground truth. We excluded any \textit{edit diffs} we used in this stage from our future evaluations. We now describe the rationale for each of our individual rule categories below. 

\begin{table}[t]
    \centering
    \begin{tabular}{|l|l|}
        \hline
        Category & Rule  \\
        \hline
        Citations & is\_citation\_inserted \\
        \hline
        \multirow{5}{*}{Point-of-view} & (para\_changes == 1) AND\\
        & (comment\_matches "POV"|"pointy") AND\\
        & NOT (is\_citation\_inserted) or deleted AND \\
        & NOT (is\_template\_inserted or deleted) AND \\
        & NOT (is\_wikilink\_inserted or deleted) AND\\
        & NOT (is\_infobox\_inserted or deleted) AND\\
        & NOT (is\_multiline\_inserted or deleted)\\
        
        \hline
        \multirow{5}{*}{Clarification} & (inserted\_length\_words b/w [0,10]) AND \\
        & (deleted\_length\_words b/w [0,5]) AND \\
        & NOT (is\_citation\_inserted\_or\_deleted) AND \\
        & NOT (is\_template\_inserted\_or\_deleted) AND \\
        & NOT (is\_wikilink\_inserted\_or\_deleted) AND \\
        & NOT (is\_infobox\_inserted or deleted) AND \\
        & NOT (is\_multiline\_inserted or deleted)\\
        \hline
    \end{tabular}
    \caption{Rules for citations, point-of-view and clarification edits}
    \label{tab:rules-cat}
\end{table}

\begin{table}[t]
    \centering
    \begin{tabular}{|l|l|p{2.5cm}|}
        \hline
        Rule & Regex & Match on\\
        \hline
        is\_citation\_inserted & <ref>|\{\{Cite\}\} & segment\_inserted\\
        \hline
        is\_citation\_inserted\_or\_deleted & <ref>|\{\{Cite\}\} & both\\
        \hline
        is\_template\_inserted\_or\_deleted & \{\{[\^{}\{]+\}\} & both\\
        \hline
        is\_wikilink\_inserted\_or\_deleted & [[\^{}[+]] & both \\
        \hline
        is\_infobox\_inserted\_or\_deleted & \^{}$|$[a-zA-Z0-9 ]+=  & both\\
        \hline
        is\_multiline\_inserted\_or\_deleted & \textbackslash n& both\\
        \hline
        comment\_matches & "pov|pointy" & edit\_comment \\
        \hline
        para\_changes == 1 & len(paragraphs) == 1 & both \\
        \hline
        inserted\_length\_words b/w [0,10] & len(segment\_inserted.split()) & segment\_inserted \\
        \hline
        deleted\_length\_words b/w [0,5]  & len(segment\_deleted.split()) & segment\_deleted\\
        \hline
    \end{tabular}
    \caption{Regular expressions for the rules used in citations, point-of-view and clarification edits.\\ "Match on" specifies the portion of the edit diff on which the regular expression is applied. "Both" specifies segment\_inserted and segment\_deleted. See Figure \ref{fig:segment-example} for reference.}
    \label{tab:rules-regex}
\end{table}

\subsubsection{Citations}
Citations on Wikipedia are added using the \textit{<ref>} or \textit{\{\{Cite\}\}} tags. Note that here we focus on cases when a citation is needed, so we do not consider modifications that editors make to existing citations (e.g., changing reference URL) or when they remove existing citations. Thus, we implement our \textit{is\_citation\_inserted} rule in Table \ref{tab:rules-cat} as a regular expression that looks for complete additions of the two tags in each sentence (Table \ref{tab:rules-regex}).

\subsubsection{Point-of-View}
% Wikipedia's \npov policy \cite{WikipediaNPOV2020} is a very broad policy covering a variety of cases of bias in text. To fix bias in content, two major types of point-of-view edits happen on Wikipedia \nikola{[REF]}: 1) inline point-of-view edits \footnote{An example of inline point-of-view edit -\href{https://en.wikipedia.org/wiki/?diff=745183020}{https://en.wikipedia.org/wiki/?diff=745183020}}, where an editor deletes or replaces one or two bias words in the sentence with a more neutral word or phrase, 2) full deletion of paragraphs\footnote{An example of full paragraph deletion -  \href{https://en.wikipedia.org/wiki/?diff=745912558}{https://en.wikipedia.org/wiki/?diff=745912558}}, or sentences not following the WP:NPOV policy\cite{WikipediaNPOV2020}. To illustrate our method, we focus on inline point-of-view edits.

Wikipedia's \npov policy \cite{WikipediaNPOV2020} is a very broad policy covering a variety of cases of bias in text. Wikipedia editors may choose to remove bias in the content inline\footnote{An example of inline point-of-view edit -\href{https://en.wikipedia.org/wiki/?diff=745183020}{https://en.wikipedia.org/wiki/?diff=745183020}}, if a few words in the paragraph are violating the neutral point-of-view (NPOV) policy, by removing or rephrasing the violating words. They may also remove an entire paragraph if the paragraph is written in a manner violating the NPOV policy\footnote{An example of full paragraph deletion -  \href{https://en.wikipedia.org/wiki/?diff=745912558}{https://en.wikipedia.org/wiki/?diff=745912558}}. To illustrate our method, we focus on inline point-of-view edits, where words in a single line violating the NPOV policy are removed or rephrased.

We implemented our rules as a regular expression that matches words "POV" and "pointy" in the edit comment.
We only consider \textit{edit diffs} that contain only a single changed \textit{line} to reduce the uncertainty of which changed \textit{line} the editor referred to in their comment. We skip sentences in the changed \textit{line} where all \textit{segments} contain only Wikipedia markup (e.g., citations, templates, links).

\subsubsection{Clarifications}
%Explaining clarifications - In Wikipedia, 
For the clarification category, we write regular expression rules to only consider segments with insertions between 0 to 10 words and deletions 0 to 5 words. We skip sentences in the changed \textit{line} where all \textit{segments} contain only Wikipedia markup (e.g., citations, templates, links). We also skip addition of new sentences because this often indicates adding new information (which is part of elaboration category).

\subsubsection{Using Semantic Intent Rules to Assign Positive Example Labels}
\label{sec:poseg}
\begin{figure}[h]
    \centering
    \includegraphics[trim=40 250 60 30,clip,width=\textwidth]{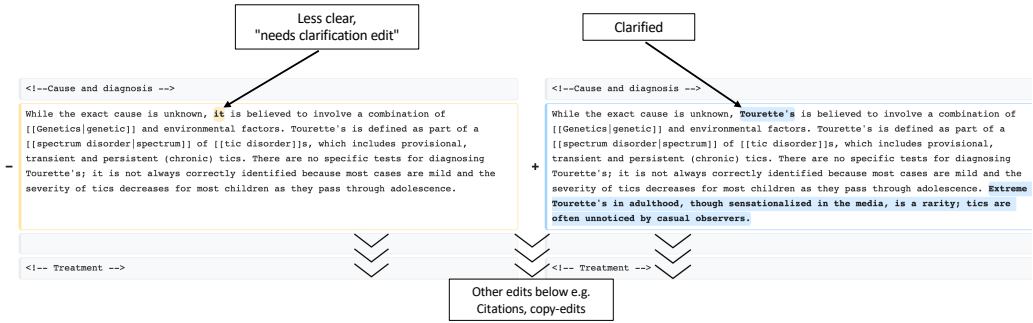}
    \caption{An edit having one of the intents as clarification and the clarified sentence. Note that the edit also contains other changes but for the purpose of "clarification", we discard the rest of the changes, that are not caught by the rules, as irrelevant. }
    \label{fig:intent-clarification}
\end{figure}

For each \textit{edit diff} that our rules detected as a specific semantic improvement category (e.g., added citation, point-of-view improvement, clarifications), we weakly label the \textit{original} Wikipedia sentence in the \textit{edit diff} with the corresponding label (e.g., needs a citation, needs point-of-view improvement, needs clarifying). For example, Figure \ref{fig:intent-clarification} shows an \textit{edit diff}, whose semantic intent is clarification, alongwith the original sentence that the editor clarified, resulting in the new clarified sentence. The original sentence is \textit{"While the exact cause is unknown, it is believed to involve a combination of [[Genetics|genetic]] and environmental factors."}. This sentence is a positive example for "needs clarification" because it was clarified to \textit{"While the exact cause is unknown, \underline{Tourette's} is believed to involve a combination of [[Genetics|genetic]] and environmental factors."}

\subsection{Leveraging Article Quality Labels to Extract Negative Examples}
\label{sec:negeg}
To extract sentences that do not need improvement (i.e., negative examples) for each semantic category, we extract sentences from highest quality Wikipedia articles (i.e., Featured Articles). Sentences in Featured Articles have sparse issues as they have gone through an extensive review process. Our insight is that it is unlikely for sentences needing improvements by Wikipedia standards to make their way into "Featured Articles". This approach of using Featured Article sentences as negative examples has been explored for detecting citations~\cite{DBLP:conf/www/RediFMT19}, where every sentence without a citation is a negative example (i.e., a sentence that does not need a citation). We generalize this in our work to extract negative examples corresponding to point-of-view and clarifications and assume that Wikipedia sentences in such articles do not need further changes to point-of-view or further clarifications.

\section{Evaluation of Semantic Intent Rules}
\label{sec:evalrules}

%\subsection{Study Design}
%\nikola{This is tasks and procedures. Method is more along the lines of what kind of study is this, comparative, what are we trying to get out of it.}
Here, we obtain ground truth semantic intent labels (citations, point-of-view, and clarifications) for a random sample of Wikipedia edits and compare them with the output of our semantic intent rules to evaluate our automatic Wikipedia labeling approach. We obtain the ground truth labels from Wikipedia editors. The purpose of this study is two-fold: 1) to assess the effectiveness of our rule-based method compared to ground truth, and 2) to understand to what extent Wikipedia editors agree on such labels among themselves.

The effectiveness of our labeling approach relies on two key factors: 1) there are more than a billion Wikipedia edits to extract positive examples from, and 2) there are enough curated Featured Articles to extract negative examples from. Even if our rules have a low recall (i.e., they miss to extract sentences that need improvement), the enormous amount of edits ensures that we are still left with a large amount of data to train deep learning models on, as long as our rules have high precision (i.e., they do not wrongly mark sentences that do not need improvement as positive examples). Note that a low recall of our rules that we use to extract positive examples would not necessarily impact the quality of our labels because our negative examples (i.e., sentences that do not need improvement) come from Featured Articles.
%Thus, to validate that our rule-based method, we measure the precision and recall of our semantic category labels.

\subsection{Study Software}
\begin{figure}[h]
\centering
\begin{subfigure}{0.5\textwidth}
  \centering
  \includegraphics[trim=60 650 300 30,clip,width=\textwidth]{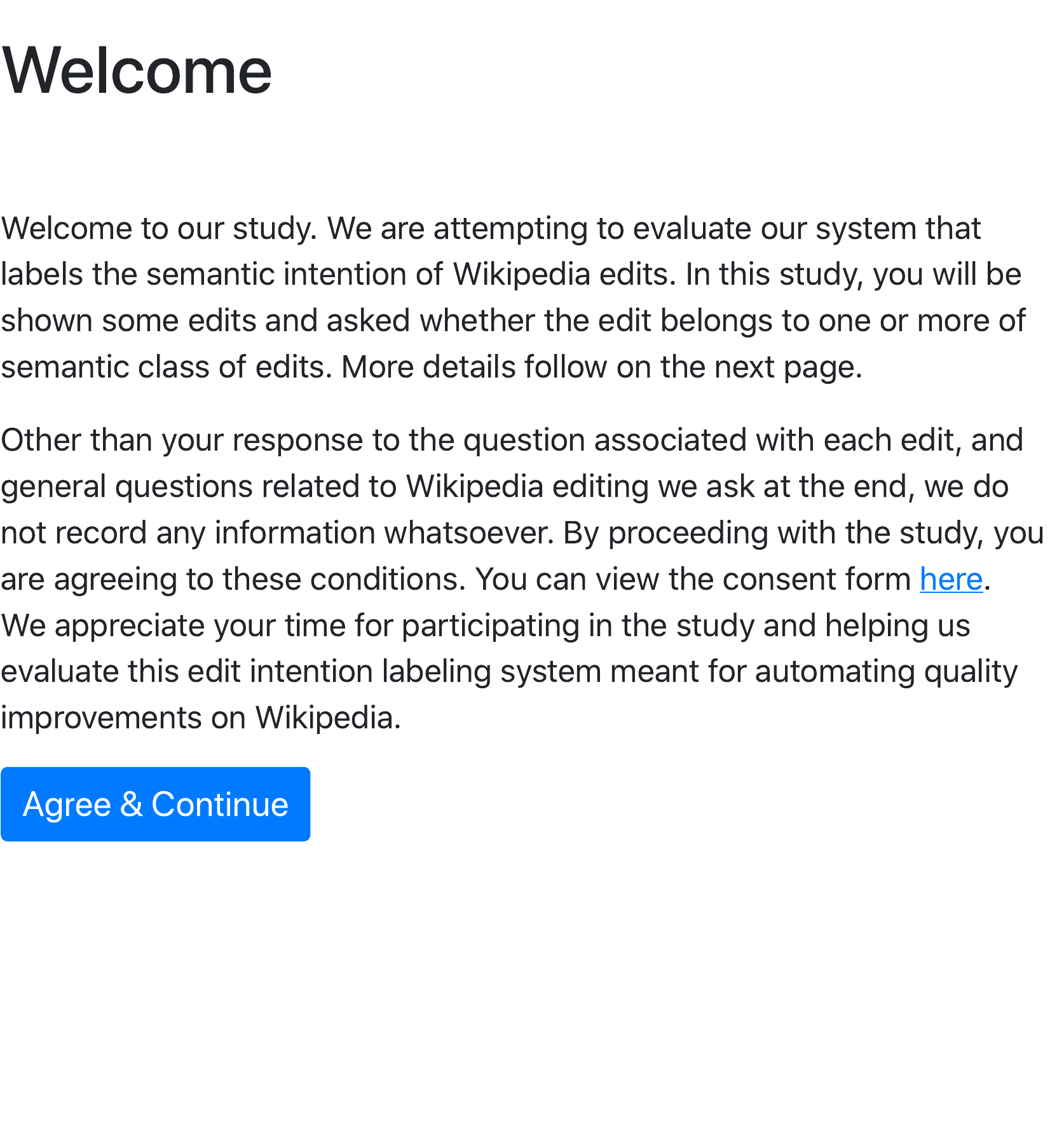}
  \caption{Welcome page}
  \label{fig:study_welcome}
\end{subfigure}%
\begin{subfigure}{0.5\textwidth}
  \centering
  \includegraphics[trim=40 550 220 30,clip,width=\textwidth]{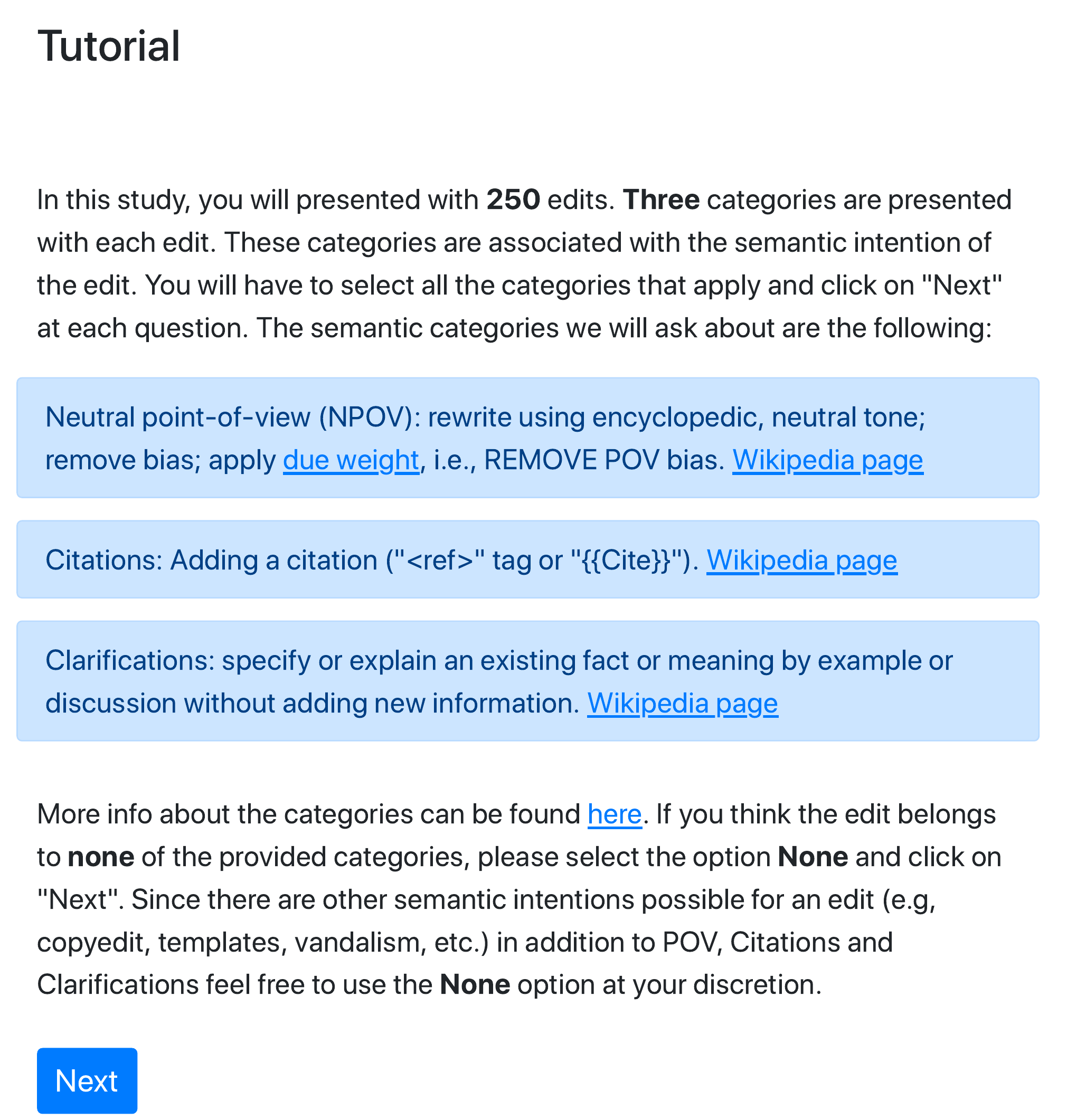}
  \caption{Definitions page}
  \label{fig:study_def}
\end{subfigure}
\caption{User study software welcome and the definitions pages.}
\label{fig:userstudy}
\end{figure}

\begin{figure}[h]
\centering
    \includegraphics[trim=40 640 40 30,clip,width=0.8\textwidth]{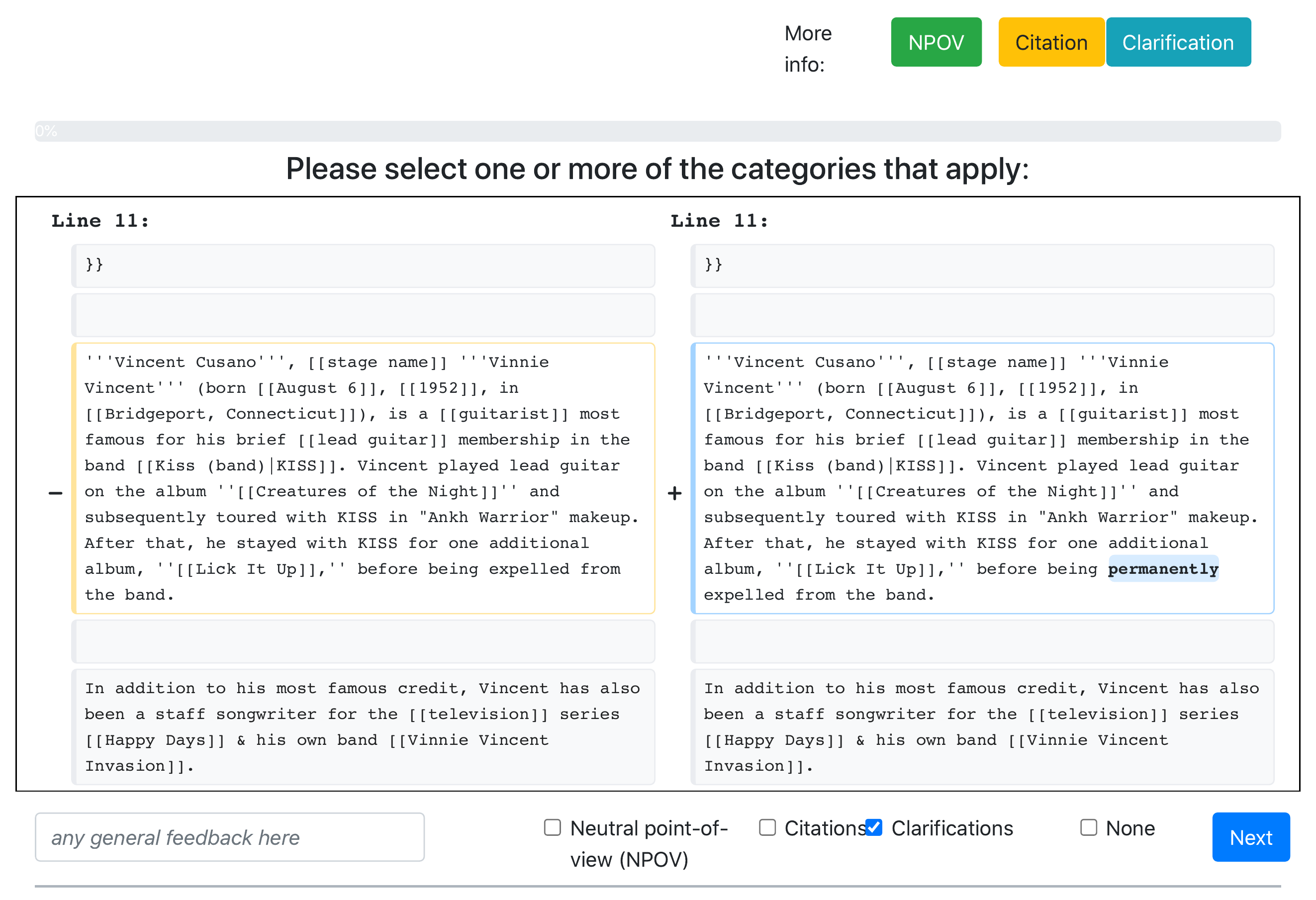}
  \caption{User study interface for labeling semantics of a Wikipedia \textit{edit diff}.}
  \label{fig:study_edits}
\end{figure}

To conduct our study, we built an online labeling interface for labeling semantic intention of different Wikipedia edits (Figure \ref{fig:userstudy} and Figure \ref{fig:study_edits}). The interface first shows the welcome page which contains a study description, a link to the consent form, and a button to consent and continue (Figure \ref{fig:study_welcome}). After the study participant clicks on the button, the interface shows the tutorial page (Figure \ref{fig:study_def}), which contains the instructions for labeling the edits using our interface along with the exact Wikipedia definitions of semantic edit intentions \cite{WikipediaEdittypesTaxonomy2020} for the three types of edits we used in our study: 1) citations, 2) point-of-view (POV), and 3) clarifications.

To label edits, our labeling interface shows one Wikipedia \textit{edit diff} at a time, similar to the Wikipedia interface (Figure \ref{fig:study_edits}). The participant can label each edit by selecting checkboxes next to the three semantic intention labels (\textit{citations, point-of-view, clarifications}), or selecting \textit{None} (if none of those three labels apply), adding optional comment, and clicking on the submit button. We removed metadata information (edit comment, author, and the date of edit) to remove any source of bias or leaking of labels. The interface always displays one practice edit to get familiar with our labeling interface.

% \begin{figure}[h]
% \centering
% \begin{subfigure}{0.3\textwidth}
%   \centering
%   \includegraphics[width=1\linewidth]{images/study1.png}
%   \caption{Welcome page}
%   \label{fig:study_welcome}
% \end{subfigure}%
% \begin{subfigure}{0.33\textwidth}
%   \centering
%   \includegraphics[width=1\linewidth]{images/study2.png}
%   \caption{Definitions page}
%   \label{fig:study_def}
% \end{subfigure}
% \begin{subfigure}{0.26\textwidth}
%   \centering
%   \includegraphics[width=1\linewidth]{images/study3.png}
%   \caption{Edit labeling interface}
%   \label{fig:study_edits}
% \end{subfigure}
% \caption{Snapshots of our study interface}
% \label{fig:userstudy}
% \end{figure}

% \begin{figure}[h!]
% \centering
%     \includegraphics[trim=60 630 300 30,clip,width=0.8\textwidth]{images/Welcome.pdf}
%   \caption{Welcome page}
%   \label{fig:welcomien}
% \end{figure}

% \begin{figure}[h!]
% \centering
%     \includegraphics[trim=40 550 220 30,clip,width=0.5\textwidth]{images/description.pdf}
%   \caption{Welcome page}
%   \label{fig:tutorial}
% \end{figure}

%3 assessments per edit. 

\subsection{Edit Diff Sampling Method}
Just like it is costly and time consuming to get training labels manually, it is similarly costly and time consuming to get ground truth labels to validate our rules. However, we can still evaluate the effectiveness of our rules on much smaller samples than it would take to train a large model. Therefore, we only sample a small random set of \textit{edit diffs} that Wikipedia experts can manually examine and label in a reasonable amount of time. Through pilot studies, we determined that 250 \textit{edit diffs} is a reasonable upper bound that a Wikipedia editor can label in an hour.

We wanted to ensure a balanced representation of \textit{edit diffs} with different semantic intents, but also have enough representative samples to properly estimate the false positive rate (i.e., the percentage of \textit{edit diffs} that our rules wrongly label as positive examples). We hypothesized that the percentage of edits with some semantic intent categories would be small (e.g., point-of-view) compared to more common ones (e.g., copy editing, wikification), and that we would risk not including them in our sample if we simply randomly selected a small subset of edits from over one billion Wikipedia edits.
%Thus, to have a balanced percentage of \textit{edit diffs} with different semantic intents (i.e, to stratify our sample), we had to first get a sense of the proportions of citations, point-of-view, and clarifications in all Wikipedia edits.
%We started with a random sample of about 100,000 edits and removed all edits with semantic intent categories that are trivial to detect using regular expressions (e.g., templates, links, metadata) or edits that we do not consider in our study (e.g., \textit{edit diffs} with changes that span multiple lines). Two authors then manually labeled a random sample of 150 edits each (300 total). We found that proportion of citation edits were approximately 15\%, point-of-view were approximately 2\%, and clarifications were approximately 5\%.
Thus, we started with a random sample of about 100,000 \textit{edit diffs} and pre-labeled them using our rules.
We then created a stratified sample of 1,000 \textit{edit diffs} by randomly selecting (without replacement) 100 pre-labeled point-of-view and clarification \textit{edit diffs} each, and then randomly selecting another 800 \textit{edit diffs} from the remaining \textit{edit diffs} without replacement.

\subsection{Participants}
We recruited nine English Wikipedia editors as participants in our study. We recruited the participants by sending emails to Wikipedia editors (using the "mail" feature on their Wikipedia user pages) who frequently discuss articles on the "Featured Article Criteria" (FAC) discussion board \cite{WikipediaFeatured2020} on the English Wikipedia. This discussion board is used for promoting articles to the "Featured Article" status; editors engaging on this forum are expected to have a good knowledge of Wikipedia editing and Wikipedia quality criteria. The average experience of the participants was approximately 10 years. The topic interests of the editors were varied including but not limited to: military history, medicine, weather, gaming, politics. Two participants were coordinators on individual Wikiprojects, one participant had an administrator role, and another participant was a new page patroller. The remaining five indicated they did not have any specific roles on Wikipedia. We compensated participants 30\$ per hour of their time they spent participating in our study.

\subsection{Tasks and Procedures}
One of the authors acted as the investigator and conducted the study with each participant separately using video conferencing. The investigator briefed each participant about the study and then provided the participants with a link to our study labeling interface (Figure \ref{fig:userstudy}). Only participants that read the consent form, which was approved by our Institutional Review Board (IRB), and consented were allowed to proceed with the study.

The participants then read the tutorial (Figure \ref{fig:study_def}) and labeled one practice edit to get familiar with our labeling interface (Figure \ref{fig:study_edits}). This was followed by an \textit{edit diff} labeling session where the interface asked participants to label the semantic intent category of 250 Wikipedia \textit{edit diffs} one at a time. The investigator asked the participants to stop labeling when the hour was up or when they labeled 250 \textit{edit diffs}, whichever came first. The participants each on average labeled 130 \textit{edit diffs} (min=45, max=250). At the end of the study, the investigator asked the participants about their Wikipedia editing experience.

\subsection{Results}
The participants labeled a total of 434 out of 1,000 \textit{edit diffs} in our sample. We ensured that each of the 434 \textit{edit diffs} had labels from at least three different participants. We had multiple participants label the same \textit{edit diffs} because point-of-view and clarifications are highly subjective categories, and a single participant may easily miss them. Thus, we assign a ground truth category to an \textit{edit diff} if at least one participant labeled the \textit{edit diff} with that category.

\subsubsection{Participant Agreement}
We first investigated the quality and ambiguities of the ground truth labels. The proportion of ground truth citations, point-of-view, and clarifications were: 24\%, 41\%,  and 75\% respectively. Note that they do not sum up to 100\% because an \textit{edit diff} can be in multiple categories. We then computed the agreement between different participants and their labels for the three categories using the Krippendorf alpha~\cite{klaus1980content}, which measures the inter-annotator agreement with more than two raters. The Krippendorf alpha~\cite{klaus1980content} values for citations, point-of-view, and clarifications were: 0.59, 0.02, 0.17, indicating medium agreement for citations, but very low or no agreement for the other two categories.

We interpret that the high agreement of citation labels was likely due participants' common grounding about what constitutes adding a citation and what does not. However, some of the disagreement could have been due to a few participants who labeled some \textit{edit diffs} as "citations added" even when an existing citation was modified, but no new citations were added.

Point-of-view and clarification ground truth labels had very high disagreement which could stem from the inherit ambiguity of the neutral point-of-view (NPOV) policy~\cite{WikipediaNPOV2020} and the clarification guidelines~\cite{WikipediaClarify2020}, and the differences in how different editors understand and interpret them. Three participants also mentioned that some \textit{edit diffs} required specific knowledge on the topic of the article to determine if the changes were addressing point-of-view.

Previous work showed such disagreements in part-of-speech tagging~\cite{plank2014linguistically}, disagreements because of lack of labelers' expertise in the area~\cite{lee2018exploring, sen2015turkers}, and disagreements on controversial matters~\cite{hickman2020wiki} (e.g., use of "annexation" \textit{vs} "liberation" when describing a conflict based on the political and ideological stand of the editor). Disagreements on the question of neutrality is a reflection of the pluralist voices of a society~\cite{10.2307/285456}. Thus, we expect that such disagreement will also impact the effectiveness of our automated rules. In Section \ref{sec:disc}, we discuss how learning from spaces with such disagreements could provide a way to probe existing policies of a socio-technical system.

\subsubsection{Effectiveness of Automated Rules}
\begin{figure}[h]
    \includegraphics[width=0.8\textwidth]{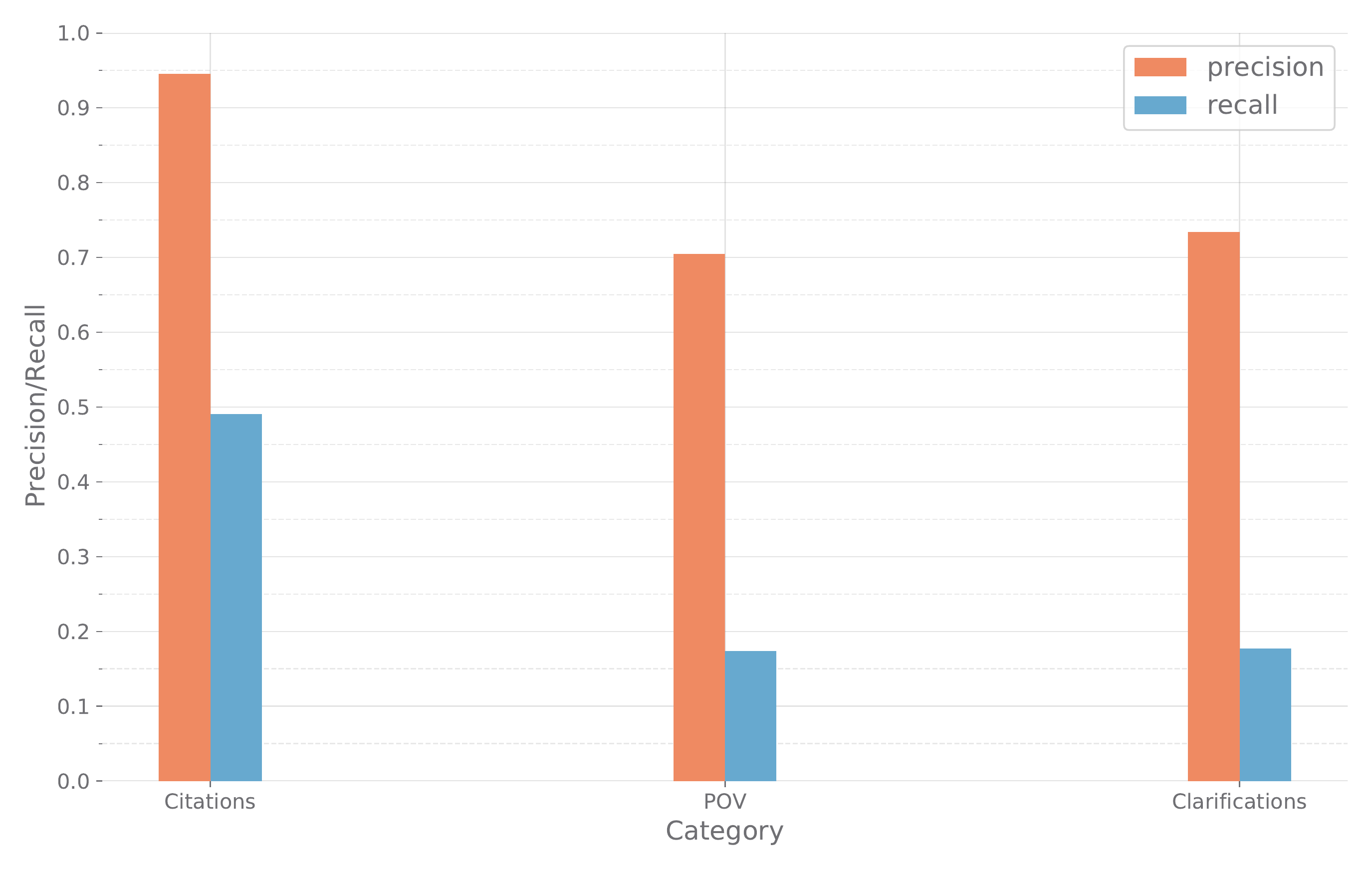}
  \caption{Precision and Recall of our rules.}
  \label{fig:prec_rec}
\end{figure}

We then evaluated the effectiveness of our rules. The proportion of citations, point-of-view, and clarifications assigned by our rules were: 12\%, 10\%, and 18\%. We computed the precision (i.e., how many of the \textit{edit diffs} that our rules extracted were actually positive examples) and recall (i.e., how many of the \textit{edit diffs} positive examples did our rules extract) of our automated labels compared to the ground truth (Figure \ref{fig:prec_rec}). The precision was 94\%, 70\%, and 73\% for citations, point-of-view, and clarification labels respectively. The recall was 49\%, 17\%, and 17\% for citations, point-of-view, and clarifications respectively.

Our results show that our citation labels had high precision. This means that our rules are able to extract positive examples that are free from false positives (i.e., sentences that our rules indicate need improvement, but that do not actually need that specific improvement category). Although it is possible that some participants confused modifying citations for adding citations (as we noted above), note that this only affected the recall (made it lower), which does not have an impact on the ability of our rules to extract positive examples.

The precision of our point-of-view and clarification labels was not as high as with citations. However, considering that our point-of-view rules use comments from editors that explicitly indicate the semantic intent behind their \textit{edit diff} (i.e., they were positive examples based on judgement from at least one editor), it is likely that our participants simply missed to properly label them or that they disagreed due to inherit ambiguity of point-of-view \textit{edit diffs}.

Similarly, clarifications can also be ambiguous. However, since our clarification rules do not use explicit labels like point-of-view rules do, we had to perform additional manual error analysis to understand the disagreement between our rules and the ground truth. Thus, we performed visual examination of false positive \textit{edit diffs} (i.e., \textit{edit diffs} that our rules extracted as positive examples, but that our participants did not label as such). We found that most false positive clarification \textit{edit diffs} could easily be confused for copy editing (i.e., adding small pieces of information), which often does add clarity to the edited sentence. In a few cases, the original editor indicated that they attempted to clarify content in the \textit{edit diff} comment.

Nevertheless, the precision for all three categories is still encouraging. The low recall presents no concerns as long as we are able to extract enough positive examples using our rules-based method. Thus, we next evaluate our ability to extract labels using our semantic intent rules.

% or expressing the confusion "could be a clarification or could be a point-of-view". Point-of-view issues sometimes involved removing or adding a word to make the statement more neutral and got confused with clarifications. The participants also assumed some grammar fixes as clarifications.

% Any work attempting to aid in addressing point-of-view issues needs to be aware of such disagreements or have a human in the loop.

% This points to the fact of having more actionable definitions for semantic intents, for both labeling and modeling  \nikola{Really? Maybe it is simply another ambiguous concept?}. For example, assessing that an edit is clarifying a location or a person's profession is less ambiguous than assessing an open ended clarification definition  \nikola{What do you mean?}. \nikola{What are the implications?}

\section{Evaluation of Sentence Quality Labels}
\label{sec:evalmodel}
We evaluate the effectiveness of our automatically generated sentence quality labels, by training Machine Learning models with our labeled sentences to perform the task of classifying whether a Wikipedia sentence needs a specific semantic improvement or not. We will refer to sentence quality labels generated using our approach as \textit{Edit-labels} henceforth.

We compare the effectiveness of our labels on classification of two semantic improvement tasks that existing research has attempted previously: 1) citations--given a Wikipedia sentence, identify whether it needs citations or not~\cite{DBLP:conf/www/RediFMT19}, and 2) point-of-view--given a Wikipedia sentence, identify whether it is biased or not according to Wikipedia's NPOV policy~\cite{DBLP:conf/wsdm/HubeF19}. For the semantic category clarification, we do not have any prior work to compare against, but we include it to showcase a category that would have been challenging (or even impossible) to detect using the existing automated labeling methods. We provide our own analysis for this category and the interpretations of the model output.

\subsection{Wikipedia Sentence Sampling Method}
We extracted 6.5 million Wikipedia \textit{edit diffs} from a random sample of 100,000 Wikipedia articles with quality ranging from the most basic "stub" articles to "Featured Articles". We filtered out \textit{edit diffs} that were reverts~\cite{DBLP:conf/wikis/EkstrandR09}, because reverted changes are not part of any of our semantic intent categories.
%reverted edits did not make it to the contents of the article.
We used the \textit{mwreverts} library\footnote{https://pythonhosted.org/mwreverts/} to label edits that were reverted and exclude them from the labeling step. We used a revert window of 15 edits and a maximum revert time of 2 days to identify \textit{edit diffs} that were reverted~\cite{DBLP:conf/ht/FlockVS12}.

We then labeled our set of non-reverted \textit{edit diffs} with their semantic intent using our rules to get positive examples as we describe it in Section \ref{sec:poseg} . We extracted the negative examples from "Featured Articles" in our sample as described in Section \ref{sec:negeg}. Table \ref{tab:dataset} shows the record counts in our final dataset for each of the three semantic categories: 1) citations, 2) point-of-view, and 3) clarifications. We further split our dataset into training (70\%), validation (10\%), and test (20\%) sets.

\begin{table}[h]
    \centering
    \begin{tabular}{|l|p{1.5cm}|p{1.5cm}|p{1.5cm}|p{1.5cm}|}
        \hline
        Dataset &  Train & Validation & Test & Total\\
        \hline
        Citations & 68,000 & 9,780 & 19,561 & 97,341 \\
        Point-of-view & 129,500 & 18,500 & 37,000 & 185,000\\
        Clarifications & 139,608 & 19,944 & 39,888 & 199,440\\
        \hline
    \end{tabular}
    \caption{Dataset splits for citations, point-of-view, and clarification categories created from edit-labels.}
    \label{tab:dataset}
\end{table}

\subsection{Model Training and Wikipedia Sentence Representation}
In our evaluation, we used the same Recurrent Neural Network (RNN) models from previous work~\cite{DBLP:conf/www/RediFMT19, DBLP:conf/wsdm/HubeF19} and only varied the labeled datasets they were trained on. We used GRU-based RNNs with global attention~\cite{DBLP:conf/emnlp/ChoMGBBSB14, DBLP:journals/corr/BahdanauCB14}, and implemented the models using tensorflow\footnote{https://www.tensorflow.org/} with keras\footnote{https://keras.io/}. The input to these models is the sentence represented as a sequence of numerical features, one for each word. When training the models, we strip the input sentences of all the wiki-markup, remove special characters and convert all text to lowercase. We used the following input features:  

\subsubsection{Word Representation}
First part of sentence representation consists of word embeddings $(w_1, w_2...w_n)$. We used GloVe word embeddings \cite{DBLP:conf/emnlp/PenningtonSM14} to represent each word. The dimensions for each word embedding were $W_{emb} \in \mathbb{R}^{100}$.

\subsubsection{Part of Speech Representation}
In addition to word embeddings described above, we used the sequence of Part-of-speech (POS) tags (one for each word) $(p_1, p_2, ...p_n)$ as input features. Like the word feature, each POS tag is represented in the 100-dimensional embedding space $W_{pos} \in \mathbb{R}^{100}$. POS tags~\cite{biber1991variation} have found extensive use in the NLP community because of their usefulness in capturing additional information about relations between words. Unlike GloVe word embeddings which were pre-trained, we trained the pos-tag embeddings with the classification task.
We used POS tag representation for training only \textit{point-of-view} and \textit{clarification} detection models because baseline work on citations~\cite{DBLP:conf/www/RediFMT19} did not use POS tags as input.

\subsubsection{Section representation}
Previous work on detecting sentences that need citations~\cite{DBLP:conf/www/RediFMT19} has shown that adding Wikipedia section representation along with word representations improved the model performance. This is because on English Wikipedia, different sections have different guidelines for citations\footnote{https://en.wikipedia.org/wiki/Wikipedia:Citing\_sources}. For example, sections like \textit{History} describing historical events need more citations than a section on the plot of a movie.

We used the section inputs only for citations detection. We first made Wikipedia section titles consistent with Wikipedia database format by converting spaces to underscores and the first character in the title to upper-case using \textit{mediawiki-utilities}\footnote{https://pythonhosted.org/mediawiki-utilities/lib/title.html}. We then trained the section embedding matrix $W_S \in \mathbb{R}^{100}$ as part of the classification objective and combined it with the word embeddings.

% \subsubsection{RNN represenation}
% \begin{equation}
%     h_t = (1 - z_t) \odot h_{t-1} + z_t \odot \tilde{h_t}
% \end{equation}
% where $h_t$ and $z_t$ are computed as follows:

% \begin{align*} \label{eq:rnn}
%     z_t &= \sigma{(W_zw_t + U_zh_{t-1} + b_z)}\numberthis\\
%     \tilde{h_t} &= tanh(W_hW_t + r_t \odot (U_hh_{t-1} + b_h))\numberthis\\
%     r_t &= \sigma(W_rw_t + U_rh_{t-1} + b_r)\numberthis
% \end{align*}
% https://tex.stackexchange.com/questions/42726/align-but-show-one-equation-number-at-the-end

\subsection{Results}
Here, we report the results of evaluation of individual semantic intent categories. Note that, unlike our label extraction that favors only extracting as few false negatives (i.e., high precision), the goal for sentence quality improvement detection task is to detect as many examples that need improvements as possible (i.e., high recall) without selecting too many false negatives (causing low precision). Thus, in our evaluation we focus on the F1-score (the harmonic mean of precision and recall) and favor models with high F1-score. However, when comparing models with similar F1-scores, having a higher recall should be preferred to detect as many sentences that need improvement.

\subsubsection{Evaluation of labels for citations}
Table \ref{tab:citation_stats} shows a comparison between the models trained on the baseline citations labeling method and our method. We tested both the model trained from the Featured Article dataset and the model trained on two datasets: 1) LQN-full, and 2) Edit-labels. We extracted the "LQN-full" dataset from the same articles as "LQN" in Redi et al.\cite{DBLP:conf/www/RediFMT19}. We used both citations and "citation-needed" tags as positive signals. Wikipedia editors add "citation-needed" tags on Wikipedia pages where they think a sentence needs a citation. Thus, in addition to actual citations, this is also a positive signal for "needing citations" because atleast one Wikipedia editor thought so. We take sentences with "citation-needed" tags or actual citations on these articles as positive examples of needing citations. We take sentences in paragraphs with no citations or citation-needed tags as negative examples. We call our dataset "LQN-full"  because we extract all the sentences from these articles giving us a large set of test sentences to evaluate citations ($\sim$300,000) compared to ($\sim$20,000) in the "LQN" dataset of the previous work. "Edit-labels test" dataset is the test split of the citation dataset created from "Edit-labels" described in Table \ref{tab:dataset}.

\begin{table}[ht]
    \centering
    \begin{tabular}{|l|l|l|l|l|l|}
        \hline
        & & \multicolumn{2}{c|}{Training} & \multicolumn{2}{c|}{Testing}\\
        \hline
        Training & Testing & Positive & Negative & Positive & Negative\\
        \hline
        Featured Articles & LQN-full & 10,000 & 10,000 & 149,000 & 151,000\\
        \hline
        Featured Articles & Edit-labels & 10,000 & 10,000 & 9,782 & 9,782\\
        \hline
        Edit-labels & LQN-full & 39,122 & 39,122 & 149,000 & 151,000\\
        \hline
        Edit-labels & Edit-labels & 39,122 & 39,122 & 9,782 & 9,782\\
        \hline
    \end{tabular}
    \caption{Statistics for the citation datasets used for training and testing.}
    \label{tab:cite_dataset_stats}
\end{table}

\begin{table}[h]
    \centering
    \begin{tabular}{|p{2cm}|p{0.5cm}|p{0.5cm}|p{0.5cm}|p{0.5cm}|p{0.5cm}|p{0.5cm}|}
        \hline
        Training & \multicolumn{6}{|c|}{Testing}\\
        \hline
        & \multicolumn{3}{c|}{\parbox[t]{1.5cm}{LQN-full\\ (300,000 samples)}} &
        \multicolumn{3}{c|}{Edit-labels test}\\
        \hline
        & P & R & F-1 & P & R & F-1\\
        \hline
        Featured Articles & \textbf{0.65} & 0.38 & 0.48 & 0.48 & 0.34 & 0.40\\
        \hline
        Edit-labels & 0.54 & \textbf{0.67} & \textbf{0.60} & \textbf{0.65} & \textbf{0.73} & \textbf{0.69}\\
        \hline
    \end{tabular}
    \caption{Testing results for citations when trained on labels from Featured Articles vs Edit-labels }
    \label{tab:citation_stats}
\end{table}

% \begin{table}[h]
%     \centering
%     \begin{tabular}{|p{2cm}|p{1.5cm}|p{1.5cm}|p{0.5cm}|p{0.5cm}|p{0.5cm}|p{0.5cm}|p{0.5cm}|p{0.5cm}|p{0.5cm}|p{0.5cm}|}
%         \hline
%         Dataset & \parbox[t]{1.5cm}{Positive\\Examples} & Negative Examples & \multicolumn{4}{c|}{"Redi et. al"} & \multicolumn{4}{c|}{Our work} \\
%         \hline
%         & & & P & R & F-1 & ROC AUC & P & R & F-1 & ROC AUC\\
%         \hline
%         LQN & 12302 & 10000 & 50 & 52 & 51 & 0 & 34 & 71 & 46 & 0\\
%         \hline
%         Edit-labels & 36448 & 32440 & 52 & 53 & 53 & 0 & 79 & 82 & 81 & 0 \\
%         \hline
%     \end{tabular}
%     \caption{Citation stats}
%     \label{tab:citation_stats}
% \end{table}

\begin{figure}
    \centering
    \includegraphics[width=0.8\textwidth]{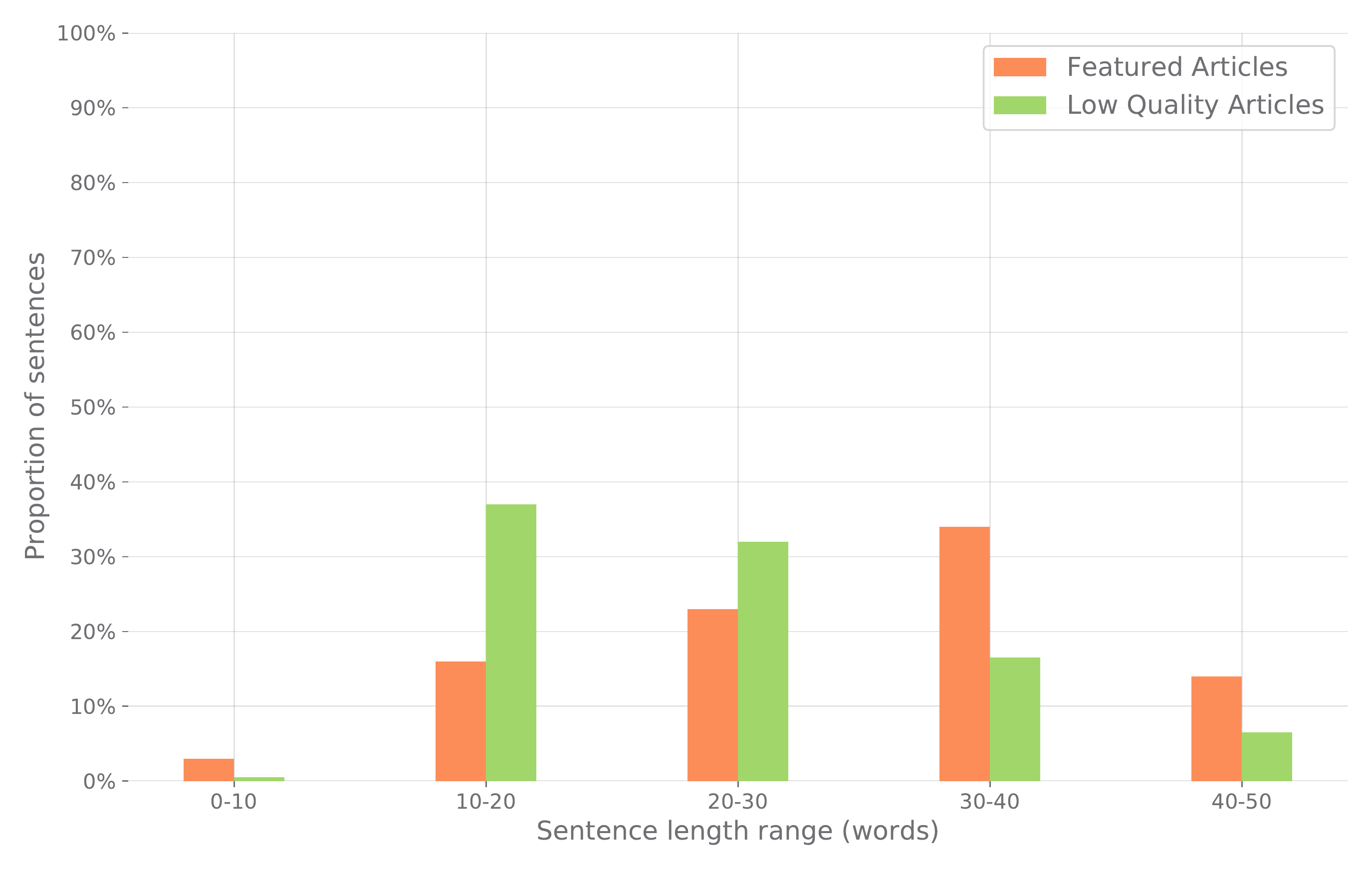}
    \caption{Distribution of sentence length (words) in Featured and low-quality articles}
    \label{fig:citation_datasets_len}
\end{figure}

For both the testing datasets, we can see that the model trained on edit-labels outperforms the model trained on limited examples from Featured Articles. We also see that when we test on a large set of sentences from low quality articles, both the approaches do not perform very well in terms of identifying citations. One of the reasons why the model trained on sentences from Featured articles does not perform very well is that the RNN model also learns other, potentially irrelevant patterns in the Featured articles (e.g., the length of the sentence). Figure \ref{fig:citation_datasets_len} shows the proportion of sentences of varying lengths (in words) in Featured Articles and low-quality articles from our datasets.

Majority of Featured Article sentences lie in the range of 20-40 words whereas low-quality article sentences have lengths in the range 10-30 words. This is expected because Wikipedia editors perform an extensive review of Featured Article sentences and ensure that they are well formed\footnote{https://en.wikipedia.org/wiki/Wikipedia:Featured\_article\_criteria}. Sentences in low-quality articles may just state a fact, not necessarily conveying the information in an elegant manner. Training on edit-labels captures this variation because we are training on sentences across the entire encyclopedia. We see that models trained on a large set of sentences extracted from edits across articles of all quality are better at generalizing to the task of identifying the need for citations.

%We re-iterate our goal here - to show that using edit labels for building models is better because they are coming from the same set of articles where the models predictions will actually be used. Improving the model classification accuracy is future work.

\subsubsection{Evaluation of labels for point-of-view}

\begin{table}[h]
    \centering
    \begin{tabular}{|l|l|l|l|l|l|}
        \hline
        & & \multicolumn{2}{c|}{Training} & \multicolumn{2}{c|}{Testing}\\
        \hline
        Training & Testing & Positive & Negative & Positive & Negative\\
        \hline
        Crowdsourced & Crowdsourced & 1,290 & 1,290 & 370 & 370\\
        \hline
        Edit-labels & Crowdsourced & 74,000 & 74,000 & 1,843 & 1,843\\
        \hline
        Edit-labels & Edit-labels & 74,000 & 74,000 & 18,500 & 18,500\\
        \hline
    \end{tabular}
    \caption{Statistics for the POV datasets used for training and testing. Crowdsourced is from previous work}
    \label{tab:pov_dataset_stats}
\end{table}

\begin{table}[h]
    \centering
    \begin{tabular}{|p{2cm}|p{0.5cm}|p{0.5cm}|p{0.5cm}|p{0.5cm}|p{0.5cm}|p{0.5cm}|p{0.5cm}|p{0.5cm}|p{0.5cm}|p{0.6cm}|}
        \hline
        Testing & \multicolumn{3}{c|}{\parbox[t]{2.5cm}{Featured \\(Crowdsourced)}} & \multicolumn{3}{c|}{\parbox[t]{2.5cm}{cw-hard \\(Crowdsourced)}} & \multicolumn{4}{c|}{Edit-labels} \\
        \hline
        Training & P & R & F-1 & P & R & F-1 & P & R & F-1 & ROC-AUC\\
        % \hline
        % featured (specific topics) & 80.1 & 79.4 & 79.7 & 62 & 85 & 72 & 91\\
        \hline
        Crowdsource trained & \textbf{0.89} & 0.71 & 0.79 & \textbf{0.71} & 0.64 & \textbf{0.67} & 0.78 & 0.67 & 0.72 & 0.70\\
        \hline
        Edit-labels trained & 0.79 & \textbf{0.85} & \textbf{0.82} & 0.42 & \textbf{0.85} & 0.56 & \textbf{0.81} & \textbf{0.86} & \textbf{0.83} & \textbf{0.92}\\
        \hline
        % featured & 89.2 & 71.5 & 79.4 & 79 & 85 & \textbf{82} & 89\\
        % \hline
        % cw-hard & 71.2 & 64.7 & 67.8 & 42 & 85 & 56 & 59\\
        % \hline
        % Edit-labels & 78 & 67 & 72 & 81 & 86 & \textbf{83} & 92\\
        % \hline
    \end{tabular}
    \caption{Testing results of Point-of-view with crowdsourcing and edit-labels training}
    \label{tab:pov_stats}
\end{table}

\begin{table}[h]
    \centering
    \begin{tabular}{|l|p{8cm}|p{3.5cm}|}
        \hline
        & Sentence & Comment\\
        \hline
        1. & After the Battle of Nicopolis in 1396 and the fall of the Vidin Tsardom three years later, the Ottomans conquered all Bulgarian lands south of the Danube, with sporadic resistance ending when the Ottomans gained a firm hold on the Balkans by conquering Constantinople in 1453. & \textbf{OK}: This looks alright to me assuming it comes with a citation that supports the firm (long term) control of the Balkans. If it's not supported by the citation, it would be POV at best.\\
        \hline
        2. & It is widely accepted that the band is joke and is collectively thought to be the biggest sale out in rock history.& \textbf{POV}. \textit{joke} is problematic\\
        \hline
        3. & Also known to have lots of friends, and it is possibly due to the fact they are known to be charming as well as loyal to friends and family members.& \textbf{POV}. \textit{charming} should be removed\\
        \hline
        4. & The family was a branch of the FitzGerald dynasty, or Geraldines, related to the Earls of Desmond (extinct), who were questionably granted extensive lands in County Limerick by the Duke of Normandy by way of conquest.  & \textbf{POV}. "Questionably" needs to be cited and it isn't phrased in a neutral tone. E.g. I would prefer "<someone> questioned the granted lands".\\
        \hline
        5. & In fact, this participation may be a reaction to the Catholic church's active political involvement. & \textbf{POV}. May be? Is? \{\{who\}\} said that?\\
        \hline
    \end{tabular}
    \caption{Manual assessment of a small sample of crowd-labeled neutral sentences}
    \label{tab:cw_hard_stats}
\end{table}
Table \ref{tab:pov_dataset_stats} shows the statistics of the datasets that we use for evaluating the labels for sentences with  point-of-view issues. The "crowdsourced" set is taken from the baseline~\cite{DBLP:conf/wsdm/HubeF19} work for point-of-view and consists of two datasets: "featured" and "cw-hard".  Hube et al.~\cite{DBLP:conf/wsdm/HubeF19} first randomly sampled 5,000 sentences from Wikipedia edit diffs that contained the comment "POV" and were single line changes. They labeled these 5,000 sentences as "biased" or "neutral" through crowdsourcing. They use the 1,843 crowdsourced labeled biased sentences from this set as positive examples in both the datasets: "featured" and "cw-hard". They used the crowdsource labeled "neutral" sentences (3,157) as negative examples in the "cw-hard" dataset. They use sentences from "Featured Articles" as negative examples for the "featured" dataset, similar to our work. 

%To compare the effectiveness of edit-labels against crowdsourced labels, we use the same GRU based attention model as Hube et. al~\cite{DBLP:conf/wsdm/HubeF19}. We only use global attention\cite{DBLP:journals/corr/BahdanauCB14} for comparison. The inputs to the model are vectors from the two sentence representations - 1) Word embeddings, and 2) Part-of-speech (POS) embeddings for the part-of-speech tags.

Table \ref{tab:pov_stats} shows testing statistics of the models when trained on the two crowdsourced datasets from previous work and our test split of the edit labels dataset. We directly report the results of the model evaluation (GRU-based RNN with global attention and word and POS tags as input) from the previous work~\cite{DBLP:conf/wsdm/HubeF19}. The model trained on our Edit-labels dataset outperforms the model trained on "featured" dataset when tested on the "featured" and Edit-labels dataset. The model did not perform at par with the baseline for the "cw-hard" dataset. The "cw-hard" dataset consists of negative examples which the crowdworkers labeled as "neutral". However, these sentences were sourced from Wikipedia edits with comment as "POV" (means atleast one Wikipedian thought there was a bias in the sentence).

To get a better sense of the crowd-labeled neutral examples from this dataset, we manually assessed some "neutral" labeled examples from the "cw-hard" dataset for clarity. One of the authors, who has researched Wikipedia editors for about a decade and is also a Wikipedia editor, identified some issues with the crowd-labeled neutral sentences in the "cw-hard" dataset. Table \ref{tab:cw_hard_stats} shows a small random sample of the negative (neutral labeled) examples from the "cw-hard" dataset with our own assessments alongwith the reasons. Four out of five sentences have point-of-view issues but they are not obvious without knowledge of the Wikipedia NPOV policies. One of the common reasons for a sentence having a point-of-view issue is not having a citation. This is because a citation pushes the opinion on the content, taking it away from the content writer which is acceptable\footnote{https://en.wikipedia.org/wiki/Wikipedia:Neutral\_point\_of\_view\#Explanation\_of\_the\_neutral\_point\_of\_view}. Crowdworkers cannot be expected to be aware of such nuances unless explicitly explained.
% \begin{table}[h]
%     \centering
%     \begin{tabular}{|p{2cm}|p{1.5cm}|p{1.5cm}|p{0.5cm}|p{0.5cm}|p{0.5cm}|}
%         \hline
%         Testing Dataset & \parbox[t]{1.5cm}{Positive\\Examples} & Negative Examples & \multicolumn{3}{c|}{}\\
%         \hline
%         & & & P & R & F-1\\
%         \hline
%         featured (specific topics) & 360 & 360 & 80.1 & 79.4 & 79.7\\
%         \hline
%         featured & 360 & 360 & 89.2 & 71.5 & 79.4\\
%         \hline
%         cw-hard & 360 & 360 & 71.2 & 64.7 & 67.8\\
%         \hline
%         Edit-labels & 19800 & 19800 & \result{78} & \result{67} & \result{72}\\
%         \hline
%     \end{tabular}
%     \caption{Testing results of Point-of-view when trained on crowdsourced labels}
%     \label{tab:pov_stats_baseline}
% \end{table}

\subsubsection{Evaluation of labels for \clarif}

\begin{table}[h]
    \centering
    \begin{tabular}{|l|l|l|l|l|}
        \hline
        Testing & \multicolumn{4}{c|}{Edit-labels}\\
        \hline
        Training & P & R & F-1 & ROC-AUC \\
        \hline
        Edit-labels & 0.75 & 0.75 & 0.75 & 0.83 \\
        \hline
    \end{tabular}
    \caption{Testing results for the clarification category}
    \label{tab:clarif_stats}
\end{table}

No prior research attempted to automatically detect whether a sentence needs clarification, hence we report and interpret our evaluation on the test set for clarifications. Table \ref{tab:clarif_stats} shows the results for clarification evaluations. We are able to achieve an F1-score of 0.75 for the clarification category and ROC-AUC of 0.83. In table \ref{tab:clarif_examples} we show some examples where clarifications were made, what clarifications were made and whether the model correctly flagged the need for the same. TP stands for true positives and FP stands for false positives. Edit-labels is used for both training and testing. Refer to Table \ref{tab:dataset} for the statistics of the training and testing splits of the Edit-labels dataset for the clarification category.

\begin{table}[h]
    \centering
    \begin{tabular}{|p{1.4cm}|p{8cm}|p{1.0cm}|p{2cm}|}
        \hline
        revision-id & Wikipedia sentence & Predi-ction & Clarification\\
        \hline
        21577990 & In sum, NLP promotes methods which are verifiable and have so far been found to be largely false, inaccurate or ineffective. & TP & ..which are \textit{largely} verifiable..\\
        \hline
        459001268 & It debuted at \#1 on the ''New York Times'' Bestseller List (''Fallen'' came in at \# 2), remaining at that position through the week of October 17. & TP & ..came in \textit{that week} at..\\
        \hline
        669670844 & During this period, Sonnenblick made the two great contributions that would define his career. & TP & ..two great \textit{scientific} contributions\\
        \hline
        N/A & In his 2013 autobiography, Jackson stated that there was, and that Martin and some white Yankees would tell racist jokes. & FP & N/A\\
        \hline
        810160688 & with  his  leg  injured  he  is  barely  able  to  get  away  but  is  rescued  by  a  kingdom  soldier  that  is  still  alive & TP & \textit{..rescued by Alavaro}, a kingdom..\\
        \hline
    \end{tabular}
    \caption{Manual assessment of clarification examples}
    \label{tab:clarif_examples}
\end{table}

\subsection{Attention}
\begin{figure}[h]
    \centering
    \includegraphics[width=0.90\textwidth]{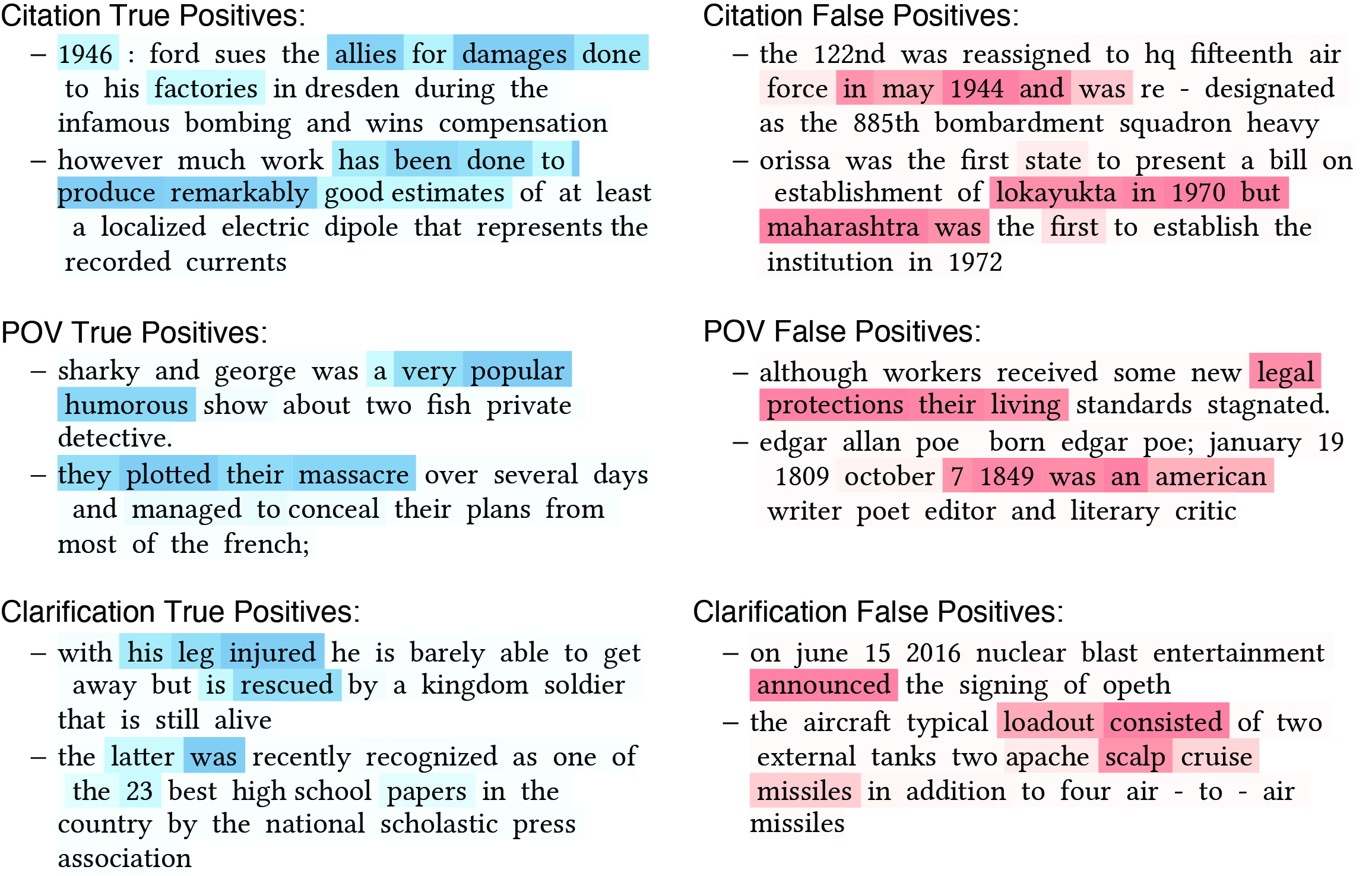}
    \caption{Attention visualization of examples for the three categories}
    \label{fig:att}
\end{figure}
Attention over RNN models~\cite{DBLP:journals/corr/BahdanauCB14} allows RNN models to place different weights on different words in the input sentence while making the predictions. We use these weights to visualize the focus that the classifier places on the different words for the given sentence. This is useful for analyzing the model outputs by the users in actual deployments. Editors can look at the most important words for a given task and take quick decisions instead of reading the full sentence. Figure \ref{fig:att} shows the attention weights for some sentences from each of citations, point-of-view and clarifications. Blue is used for true positives and red is used for false positives. The darker the word, the more the attention on that word for prediction for the specific sentence. 

For citations, attention is given to reporting verbs like \textit{damages done} or \textit{produce remarkably good estimates} which require proof of opinion. This is in line with previous work\cite{DBLP:conf/www/RediFMT19} who report that such verbs or presence of facts lead to high likelihood of needing citations. For point-of-view, attention seems to be particularly helpful as focus is given on words such as \textit{very popular}, \textit{formidable}, \textit{plotted their massacre} which are the typical deletions made in inline point-of-view edits as per the guideline - \textit{avoid stating opinion as facts} \footnote{https://en.wikipedia.org/wiki/Wikipedia:Neutral\_point\_of\_view\#Explanation\_of\_the\_neutral\_point\_of\_view}. This is because, if such words are uncited, they cause a point-of-view issue as the opinion becomes an opinion of the content writer. For clarifications, attention is placed near the words where additions were made in the sentence. For example, consider sentence 1 in clarification true positives, \textit{rescued by a Kingdom soldier} is clarified to \textit{rescued by Alavaro, a kingdom soldier} and a high attention is placed on rescued (see table \ref{tab:clarif_examples} - last example).

\section{Discussion}
\label{sec:disc}
Our Wikipedia sentence quality labeling method and detection pipeline has several implications for both Wikipedia and collaborative systems more broadly. By learning implicitly from expert behaviors, we introduce a flexible and fast labeling method. However, our labeling method is not free from inherent ambiguities of assessing content quality, which has implications not only for how we pick our labels, but also how we train our models to detect low quality content on collaborative platforms. Finally, the models that we train from implicit behaviors enable a new family of technical probes that can provide insight into how Wikipedia editors enact existing Wikipedia policies and guidelines in practice.

\subsection{Flexible, Fast, and Effective Labeling}
%Existing research~\cite{holstein2019improving} has shown that better data collection is of greater importance than model development towards building fairer ML systems. As such, 
Our method can extract more data, more quickly compared to existing methods. By demonstrating the application of our sentence quality detection pipeline on three categories of semantic edit intents (citations, point-of-view, and clarifications), we showed that our method can be used to extract order of magnitude more data than what crowd-sourcing can get. Further, we have shown that models trained from Wikipedian behavior traces outperformed those trained using existing labeling methods.

One of the benefits of our method is that it can continue to grow the size of training datasets as new Wikipedia edits become available. This means that our method could quickly adapt labels to reflect changes in editor behaviors over time, something that would render models trained on existing labels ineffective. Even if there are updates to the semantic edit categorization, our automated method only requires updates to the corresponding rules, unlike existing manual labeling methods that would require us to hire crowdworkers or Wikipedia editors to relabel the data.

We use precision and recall metrics to evaluate our labeling approach against prior work~\cite{DBLP:conf/www/RediFMT19,DBLP:conf/wsdm/HubeF19}. However, for practical use, sentence quality detection models trained using our labeling approach can be used to flag problematic sentences (e.g., needing citations, needing point-of-view edits) on Wikipedia articles based on a manually chosen threshold. If more Wikipedia editors are available to assess the predictions, a low confidence threshold of the classifier can be used to flag more examples with issues (high recall). If the Wikipedia community wants only high quality predictions to avoid false positives, predictions with a high confidence threshold can be made (high precision).

\subsection{Inherit Ambiguity of Semantic Intent Category Descriptions and Labels}
Our findings point to the inherent ambiguity of content quality labels and different semantic edit intents that we base our rules on. This was evident in the high disagreements between editors in our study. The diverse background of Wikipedia editors brings in a pluralist mindset~\cite{matei2011wikipedia}, which introduces different interpretations of what requires improvement and in turn introduces ambiguity into content quality labels. Also, some edit type categories, such as point-of-view (POV), are highly topic dependent and some editors expressed their inability to correctly judge the category of an edit because of being non-experts in that area.

However, some of the disagreements could also come from ambiguities in the Wikipedia's description of different edit type categories ~\cite{WikipediaEdittypesTaxonomy2020}. For example, our participants' comments during the study implied that the existing definition of clarifications is fairly ambiguous, which could account for some of our participants' disagreement on the clarification labels. However, their comments also indicated that the ambiguity could come from overlaps in edit type definitions. For example, clarifications are also difficult to distinguish from elaborations when information is added in the same line. One potential way to reduce some of this ambiguity could be to introduce more concrete examples for each semantic category (e.g., what needs a clarification: a missing date, missing location, or missing profession). 

Our findings also indicate that point-of-view issues have more nuance beyond "peacock terms /weasel words" that reflect opinionated language. A major cause of point-of-view issues arises due to lack of citations as lack of credibility indicates that the content is likely an opinion of the author of the content. Many point-of-view issues also arise from different ideological stands of the editors. For example, a contention could be regarding the use of the word "annexation" vs "liberation" when describing a conflict on Wikipedia, based on which ideological side the editor belongs to.

\subsection{Models Learned from Implicit Behaviors as Technical Probes for Wikipedia Norms}
Our method operationalizes Wikipedia written policy genres (e.g., policies, guidelines, and "essays"~\cite{morgan2010negotiating}) to automatically extract quality labels from Wikipedia editor behaviors that capture how editors interpret and enact current policy genres. Wikipedia written policy genres are examples of Wikipedia online community injunctive norms (i.e., shared beliefs within a group, community or culture that  "ought" to be followed), while Wikipedia editor behaviors constitute a set of descriptive norms (i.e., typical behaviors of individuals within a group, community or culture)~\cite{cialdini2004social}.

The most common role of Machine Learning (ML) algorithms that we can train using our labels is supporting and enforcing Wikipedia's norms. For example, an ML algorithm that automatically flags low quality Wikipedia sentences is a technical representation of policy. Recent work in this area of algorithmic governance~\cite{muller2013work, lovink2012critical} discussed the rising role of algorithms as norm enforcement mechanisms in Wikipedia. However, that work strictly explores enforcing Wikipedia injunctive norms. Instead, our approach of modeling Wikipedian behavior---capturing not only the explicit rules, but rather describing how the rules are applied in practice---extends this notion of governance from originating from formal consensus (top down) to a description of actual practice (bottom up).

We further envision the role of models trained on our labels as supporting resources to reflect on Wikipedia norms themselves. Much like Wikipedia "essays" that document editor practices and provide a "soft regulatory mechanism" where a formal policy is insufficiently specific to apply~\cite{morgan2010negotiating}, predictive models that describe Wikipedian behavior when referencing a formal policy afford the ability to explore probable applications of that policy in new contexts.  Like "essays", this mixing of context with formal norm interpretation has the potential to also fill a niche in the broader normative ecology of Wikipedia and to extend the interpretability and consistent application of more abstract, formal policies in the socio-technical practice of editing Wikipedia articles.

\section{Conclusion and Future work}
Our work showed how content policies of Wikipedia can be coded into programmatic rules to identify the semantic intents of Wikipedia edits, which can then be used to detect sentences with quality issues at scale. We showed that deep learning models can be trained using these sentences labeled with quality issues that can identify quality issues in new unseen Wikipedia sentences.

Future work can directly leverage our proposed method to build new sentence quality identification pipelines to flag issues in sentences along other semantic intent categories. For example, experienced editors in our study mentioned performing a lot of copy edits, which involve fixing the grammar in the articles as a lot of Wikipedia editors are not native speakers of English. Rephrasing the tone of the article to a more encyclopedic one is another commonly performed action on the articles. One editor mentioned that "articles are written in a victorian style storytelling manner and that has to be rephrased to a more encyclopedic tone".
%These models can also be used in combination to identify more nuanced issues. A very compelling argument is with citations and point-of-view. Often, if a fact is cited, it is not considered as pushing a point-of-view. This is because, according to NPOV the guidelines\cite{WikipediaNPOV2020}, opinions in the content are alright, but opinions of the curator of the content are not allowed.

Our sentence quality labeling approach could be used to assess whether Wikipedia editors improve content quality according to existing injunctive norms. This is because policies~\cite{forte2009decentralization} mediate behavior in socio-technical systems like Wikipedia and the models trained on edits of Wikipedians capture such policy guided behaviors. By providing a way to bridge the gap between descriptive and injunctive norms~\cite{cialdini2004social} future work should explore how to make normative behavior more consistent on Wikipedia. Moreover, future work should explore how our approach could give more structure to the traditional way of guiding newcomers and passing knowledge by distilling policy guided human behaviors of the community into machine learning models and show a way to use the output of the models as an aid to understand and reason about the policies.

%Datasets of parallel sentence quality improvements can also be generated using our approach. For example, rules that can label simplification edits can be used to extract large instances of complex-simple sentence pairs for text simplification tasks. This will be a direct improvement over current text simplification approaches~\cite{yatskar-etal-2010-sake} which uses sentences from corresponding articles of Simple English Wikipedia and English Wikipedia and aligns them to generate a large corpus of complex-simple sentence pairs.

%We illustrated our work on citations, point-of-view and clarifications; however, future work should explore how to create effective rules for extracting examples for other semantic intent edit categories. For example, experienced editors in our study mentioned performing a lot of copy edits, which involve fixing the grammar in the articles as a lot of Wikipedia editors are not native speakers of English. Rephrasing the tone of the article to a more encyclopedic one is another commonly performed action on the articles. One editor mentioned that "articles are written in a victorian style storytelling manner and that has to be rephrased to a more encyclopedic tone".

Our entire pipeline (labeling semantic intention of Wikipedia \textit{edit diffs} and extracting sentences from Wikipedia articles that need improvements) is language agnostic. Future work should explore how to develop effective rules for other language Wikipedias. For example, for low resource Wikipedia language editions, effective rules can be built by exploiting semantic relatedness~\cite{hassan2009cross} of concepts from low resource to high resource Wikipedia language editions. Point-of-view edits in a target language Wikipedia could be detected by starting with point-of-view edits in English Wikipedia, and finding edits in the target language Wikipedia that are semantically "close".

%Our pipeline can also be  used to personalize models according to a specific space. This space can be specific topics to address content issues that are sensitive to topic spaces, like point-of-view or specific classes of articles. For example, an article in a C-class may need more copy edits and grammar fixes than an article in A-class which may need more structuring. In such cases, building sentence quality detection models for C-class articles will benefit from learning from edits that happen in C-class articles.

%Sentence quality detection on Wikipedia articles will be followed by actually improving the quality of articles as they evolve through various quality stages. Further, directly identifying sentence quality along different semantic dimensions allows us to do fine grained analysis of how good quality articles evolve and what improvements were made by different actors over time which led to quality improvements. Using such analysis, models can be built that can be used to accelerate quality improvements in collaborative systems or assist users in the same.

\begin{acks}
This work was funded by Toyota Research Institute (TRI). We thank members of the Machine Assisted Cognition group at TRI for valuable discussions about this work. We are grateful to Jonathan Morgan from Wikimedia Foundation for insightful discussions on injunctive and disjunctive norms in online spaces, and Diyi Yang from Georgia Tech for providing us with the classifier for the semantic edit intentions dataset.  We also thank members of the CompHCI lab at the University of Michigan, Nel Escher and Divya Ramesh, for their input on ambiguity of labels in collaborative spaces, and Anindya Das Antar and Snehal Prabhudesai, for their help in testing an early prototype of the study software.
\end{acks}

\bibliographystyle{acm}
\bibliography{sample-base}

% \section*{Acknowledgement}
% We thank the sponsors of the project from TRI - grant no...
% We are grateful to James Pennebaker for providing us with the LIWC lexicon for research use. 
\end{document}